\documentclass{JHEP3}

\usepackage{amsmath,amssymb}
\usepackage{epsfig}
\usepackage{graphicx}
\usepackage{cite}

\title{
Ricci solitons, Ricci flow, and strongly coupled CFT in the Schwarzschild Unruh or Boulware vacua
}

\author{Pau Figueras \\
DAMTP, Centre for Mathematical Sciences, University of Cambridge, Wilberforce Road, Cambridge CB3 0WA, U.K.
}

\author{James Lucietti \\
School of Mathematics and Maxwell Institute for Mathematical Sciences, University of Edinburgh, King's Buildings, Edinburgh, EH9 3JZ, U.K.}

\author{Toby Wiseman \\
Theoretical Physics Group, Blackett Laboratory, Imperial College, London,  SW7 2AZ, U.K.}

\date{April 2011}

\abstract{

The elliptic Einstein-DeTurck equation may be used to numerically find Einstein metrics on Riemannian manifolds. Static Lorentzian Einstein metrics are considered by analytically continuing to Euclidean time. Ricci-DeTurck flow is a constructive algorithm to solve this equation, and is simple to implement when the solution is a stable fixed point, the only complication being that Ricci solitons may exist which are not Einstein. 
Here we extend previous work to consider the Einstein-DeTurck equation for Riemannian manifolds with boundaries, and those that continue to static Lorentzian spacetimes which are asymptotically flat, Kaluza-Klein, locally AdS or have extremal horizons. Using a maximum principle we prove that Ricci solitons do not exist in these cases and so any solution is Einstein. We also argue that  Ricci-DeTurck flow preserves these classes of manifolds.
As an example we simulate  Ricci-DeTurck flow for a manifold with asymptotics relevant for $AdS_5/CFT_4$. Our maximum principle dictates there are no soliton solutions, and we give strong numerical evidence that there exists a stable fixed point of the flow which continues to a smooth static Lorentzian Einstein metric. Our asymptotics are such that this describes the classical gravity dual relevant for the CFT on a Schwarzschild background in either the Unruh or Boulware vacua.  It determines the leading $O(N_c^2)$ part of the CFT stress tensor, which interestingly is regular on both the future and past Schwarzschild horizons.
}

\begin{document}

\section{Introduction}

Exact solutions to Einstein's equation are notoriously difficult to find and traditionally most progress has been made  under extra symmetry assumptions. However, in recent times it has become apparent that there are many interesting solutions of the Einstein equations with less symmetry than the usual examples.\footnote{The simplest example of this is five dimensional KK static vacuum black holes, where on top of the uniform black string, one can have non-uniform black strings and localised black holes.} We are likely never to know these analytically, and so if we wish to have a detailed understanding of such -- believed to exist -- solutions, one must resort to numerical techniques. In modern times there are three key areas where such questions arise: i) string theory phenomenology, where one is interested in the Riemannian geometry of the extra dimensions,  ii) black holes in higher dimensions, where there is a very complicated phase structure, even in the static case if the extra dimensions are compact, and iii) AdS-CFT, where exotic gravitational solutions of many types have gained interesting physical significance due to their relation to certain strongly coupled CFTs.

A general approach for constructing  static vacuum solutions has been outlined in \cite{KitchenHeadrickWiseman}. The static spacetime is analytically continued to a Riemannian geometry and then the elliptic Einstein-DeTurck equation is solved as a boundary value problem. 
This approach is particularly attractive because if the spacetime contains a non-extremal black hole, after analytic continuation  with suitable Euclidean time period, there is no boundary associated to the horizon location and the Riemannian manifold is smooth there.
Two numerical algorithms have been proposed to solve the Einstein-DeTurck equations: simulation of the Ricci-DeTurck flow, and the Newton method. The Ricci-DeTurck flow is the most elegant since it is geometrical, and turns out to be simple to simulate numerically. It is this method that we focus on in this paper. 
A number of previous works have computed Ricci flows numerically in a variety of related contexts \cite{Garfinkle:2003an,Headrick:2005ch,Headrick:2006ti,Holzegel:2007zz,Doran:2007zn,Headrick:2007fk,KitchenHeadrickWiseman}.

%
One aim of the current paper is to extend the discussion of the Einstein-DeTurck equation in this numerical context to allow for new types of boundary or asymptotic conditions, beyond the simple one considered in \cite{KitchenHeadrickWiseman}. We will restrict to vacuum solutions with zero or negative cosmological constant. 
We show how to impose boundary conditions for general asymptotically Euclidean, Kaluza-Klein and locally hyperbolic Riemannian manifolds which continue to give Lorentzian geometries which are asymptotically flat, Kaluza-Klein or locally AdS. 
We also show how to define boundary conditions for static Riemannian manifolds\footnote{We will refer to a `static' Riemannian geometry as a Riemannian manifold possessing a $U(1)$ or $\mathbb{R}$ isometry generated by a hypersurface orthogonal Killing field.} whose Lorentzian analytic continuations contain extremal horizons (this corresponds to an asymptotic region in Euclidean signature). 
We also consider the Einstein-DeTurck equation on general manifolds with boundaries. We discuss the boundary conditions considered by Anderson~\cite{Anderson1}, where the conformal class of the induced metric and trace of extrinsic curvature are fixed. We also examine the case where the extrinsic curvature is proportional to the induced metric.

Furthermore, since we wish to use Ricci-DeTurck flow to solve the Einstein-DeTurck equation, we have considered whether our boundary conditions are preserved under such flows. For the asymptotically Euclidean case this has been considered previously in the literature~\cite{ToddOlnyik} where it was shown that Ricci flow preserves this class on manifolds. It seems reasonable to expect that this can be generalised to the Kaluza-Klein case, although we have not considered this. The asymptotically locally hyperbolic class has recently been proven to be preserved by Ricci flow under a smoothness assumption~\cite{Bahuaud}. For static manifolds with asymptotic regions that analytically continue to Lorentzian extremal horizons, we have proved here that they are indeed preserved by Ricci-DeTurck flow;  the key idea here is that there exists a well defined general notion of a near-horizon geometry which itself solves the Einstein equations~\cite{KLR}.

There are two subtleties to using Ricci-DeTurck flow to solve the Einstein-DeTurck equation. Firstly a static solution in vacuum gravity may not be a stable fixed point of the flow.\footnote{ Recall that the stability (under the flow) of a given fixed point is determined by the non-existence of negative eigenvalues in the spectrum of the Euclidean Lichnerowicz operator for that fixed point.} This is often the case for black hole solutions due to the existence of Euclidean negative modes \cite{GPY}. In \cite{KitchenHeadrickWiseman} it was proposed that by appropriate tuning of a family of initial data one may still find solutions, although in practice the Newton method usually works best in these situations. Therefore Ricci-DeTurck flow is best applied to cases where the fixed point of interest is believed to be stable. 

A second subtlety is that there exist solutions of the Einstein-DeTurck equation that do not correspond to solutions to the Einstein equations, but instead are Ricci solitons. To address this, we show that there is a simple maximum principle which impacts on the existence of Ricci solitons. Applying this maximum principle to the cases with boundary and asymptotic conditions considered above, in fact rules out the existence of Ricci solitons. 

In order to illustrate these general ideas, we consider an interesting example of a Ricci-DeTurck flow. 
Specifically it is a flow on a five dimensional static, axisymmetric Riemannian manifold with two asymptotic regions that is smooth in the interior. One asymptotic region is 
locally hyperbolic with Schwarzschild boundary metric, and the other has a Lorentzian analytic continuation which gives an extremal horizon whose near-horizon geometry is that of standard Poincare AdS. 
By our general results we know how to define boundary conditions such that the flow preserves such a class of Riemannian manifolds. 
Furthermore,  from our non-existence result concerning Ricci solitons, we know that {\it if} the flow converges to a fixed point, it must be an Einstein metric. 
We provide strong numerical evidence that the flow {\it does} converge, and that such a fixed point with smooth interior does exist.  
Furthermore, by the above comments we expect this fixed point is {\it stable}, since no fine tuning was required to flow to it. 

The Lorentzian continuation of this new Einstein metric has a very interesting $AdS_5/CFT_4$ interpretation. 
Namely, as the regular static classical bulk geometry dual to the Lorentzian signature CFT$_4$ on a Schwarzschild black hole background in a vacuum state. 
Indeed, the bulk geometry has an asymptotic AdS boundary in the `UV' end of the geometry, with conformal boundary metric being Schwarzschild.  Furthermore, since in the bulk the geometry approaches that of the Poincare-AdS horizon,  it is conformal in the `IR' and hence is dual to the CFT in a 
(non-thermal) 
vacuum state.  
We argue that this classical solution describes the dominant bulk saddle point in $AdS_5/CFT_4$ suitable for describing the Unruh and Boulware vacua in the CFT. Since the solution has an extremal horizon in the IR, we do not expect it gives the dominant saddle point for the CFT in the Hartle-Hawking vacuum, which presumably would be of the form of a `funnel' or `droplet' discussed in \cite{Hubeny:2009ru,Hubeny:2009kz} with a non-extremal horizon in the IR with the same temperature as the boundary black hole.

A classical bulk solution allows one to compute the leading $O(N_c^2)$ part of the dual CFT stress tensor, and we do so from our numerical solution.
We believe this is the first calculation of the leading order contribution to a stress tensor at large $N_c$
 for a four dimensional strongly coupled field theory in an asymptotically flat black hole background. 
 Since our bulk solution is smooth in the interior, we find the remarkable result that this leading contribution to the stress tensor is regular on both the future and past horizons 
 even though the solution describes the dominant classical saddle point for the dual to the 
 Unruh and Boulware vacua.
Since our calculation gives only the leading $O(N_c^2)$ contribution at large $N_c$ and is at strong 't Hooft coupling, the familiar free field theory properties of the Unruh and Boulware vacua are not apparent (i.e. the former being regular only on the future horizon and the latter regular on neither future nor past horizon).

Finally, we emphasize that this example represents an optimal situation for the numerical construction of a new solution which is of interest physically, extremely simple to find just using Ricci-DeTurck flow (as it is a stable fixed point), and where one can rule out the potential complication of existence of Ricci soliton solutions. 

This paper is divided into two main parts. In the first part we extend the analysis of  \cite{KitchenHeadrickWiseman} to understand how to impose the various boundary conditions of interest for the Einstein-DeTurck equation, and for the Ricci-DeTurck flow. This begins in \S\ref{sec:maximum} where we point out a simple maximum principle for the Ricci soliton equation. Then in \S\ref{sec:elliptic} we consider various new boundary and asymptotic behaviours for the Einstein-DeTurck equation, and show that the maximum principle applied in such cases rules out the existence of soliton solutions of the Einstein-DeTurck equation. Then in \S\ref{sec:parabolic} 
we consider Ricci-DeTurck flow as a method for solving the Einstein-DeTurck equation, and discuss whether our boundary conditions are preserved by such flows, proving this is the case for static manifolds whose Lorentzian section contain extremal horizons.
In the second part,  \S\ref{sec:example}, we illustrate these general ideas by giving an example of a Ricci-DeTurck flow, and provide strong numerical evidence that it converges to an Einstein metric with both asymptotically extremal and locally hyperbolic regions. The Lorentzian continuation of this is a smooth Einstein metric which is asymptotically locally AdS with Schwarzschild boundary metric and also contains an extremal Poincare-AdS horizon.
We conclude the paper in \S\ref{sec:discussion} with a brief summary and a discussion mainly devoted to considering the AdS/CFT interpretation of our new solution.  Some of the technical details are relegated to several appendices.

\section{Einstein-DeTurck solitons}

\subsection{A maximum principle for Ricci solitons}
\label{sec:maximum}

Let  $(\mathcal{M},g)$ be a smooth (connected) $m$-dimensional Riemannian manifold which obeys the Ricci soliton equation
\begin{eqnarray}
\label{eq:soliton}
R_{\mu\nu} - \nabla_{(\mu} \xi_{\nu)} - \Lambda g_{\mu\nu} = 0
\end{eqnarray}
for some 1-form $\xi$, where $\nabla$ is the Levi-Civita connection associated to the metric $g$. We will consider a cosmological term which is either vanishing or negative (hyperbolic), so that $\Lambda \le 0$.  We term a solution $(g,\xi)$ of this equation with non-vanishing $\xi$ to be a \emph{non-trivial} soliton.

The contracted Bianchi identity implies that any solution to the soliton equation must satisfy the vector equation,
$\nabla^2 \xi_\mu + R_{\mu}^{\phantom\mu\nu} \xi_\nu = 0$.
Combining this equation with the original Ricci soliton equation \eqref{eq:soliton}, and defining the function $\phi \equiv \xi^\alpha \xi_\alpha \ge 0$, one can establish the scalar equation
\begin{eqnarray}
\label{eq:phi}
\nabla^2 \phi + \xi^\mu \partial_\mu \phi = - 2\Lambda \phi + 2( \nabla^\mu \xi^\nu ) ( \nabla_\mu \xi_\nu ) \ge 0  
\end{eqnarray}
where we have noted that both terms on the right hand side are non-negative.

Clearly for a solution to be a non-trivial soliton it is necessary (although certainly not sufficient) for the associated equation 
\begin{eqnarray}
\label{eq:condition}
\nabla^2 f + \xi^\mu \partial_\mu f \ge 0
\end{eqnarray}
for a function $f$ on the fixed background $(\mathcal{M},g,\xi)$, to have a non-vanishing solution such that $f \ge 0$.  Furthermore, the behaviour of $f$ on any boundary $\partial \mathcal{M}$, needs to be the same as that for $\phi$.
However, this equation admits a maximum principle\footnote{Hopf's maximum principle is the relevant local result and can be found in standard texts \cite{Protter}. For related global statements on Riemannian manifolds see \cite{Aubin}.} which for non-constant $f$ states: i) $f$ may attain its maximum only on the boundary $\partial \mathcal{M}$, and ii) the outer normal gradient obeys $\partial_n f > 0$ at such a maximum on $\partial \mathcal{M}$.  Thus, in particular note that if $f\geq 0$ and $f=0$ on $\partial \mathcal{M}$, statement (i) implies that $f \equiv 0$ everywhere in $\mathcal{M}$.

On a compact manifold with no boundary, the maximum principle thus shows that $f$ must in fact be a (possibly non-zero) constant, and therefore in the soliton case one learns that $\phi$ is a non-negative constant. Furthermore, if $\Lambda<0$ the r.h.s. of (\ref{eq:phi}) in fact implies $\phi \equiv 0$ (and hence the soliton is trivial), whereas if $\Lambda=0$ one has $\nabla_\mu\xi_\nu=0$ (and hence the soliton is Ricci flat).
In fact compact solitons with $\xi \ne 0$ do not exist for $\Lambda=0$ in the absence of boundaries~\cite{Bourguignon}, although that isn't apparent from our simple maximum principle argument. 

Our interest here is in both manifolds with boundaries and non-compact manifolds.  In particular,  we are interested in general manifolds with a boundary, such that either $\phi=0$ on $\partial \mathcal{M}$, or the outward normal derivative  $\partial_n \phi \le 0$ on $\partial \mathcal{M}$. We are also interested in non-compact manifolds with a number of ends, such that $\phi \to 0$ asymptotically in each end. In these cases, from above and as we argue below, the maximum principle rules out a non-zero solution to \eqref{eq:condition}, and hence non-trivial soliton solutions.

\subsection{Einstein-DeTurck equation on manifolds with boundaries and asymptotic regions}
\label{sec:elliptic}

From now on we consider the DeTurck choice of vector field $\xi$, which is constructed from the Levi-Civita connection $\Gamma$ of $g$ and a smooth reference connection $\bar{\Gamma}$,
\begin{eqnarray}
\label{eq:xi}
\xi^\mu = g^{\alpha\beta} \left( \Gamma_{\alpha\beta}^{\mu} - \bar{\Gamma}_{\alpha\beta}^{\mu} \right)   \; .
\end{eqnarray}
This was introduced by DeTurck in the context of an elliptic problem, and later used by him to argue Ricci flow is diffeomorphic to a parabolic flow (Ricci DeTurck flow)~\cite{DeTurck}.
In this case we term the Ricci soliton equation \eqref{eq:soliton} the Einstein-DeTurck equation, and unlike the usual Einstein equation, for Riemannian signature this is elliptic. Hence we expect to be able to solve it for suitable boundary conditions as a boundary value problem.
We further make the choice that the reference connection is the Levi-Civita connection of a reference metric $\bar{g}$ on $\mathcal M$. 

Before we give some examples of boundary and asymptotic conditions of interest, it is useful to introduce a general chart $(w,x^i)$ for  the metric\footnote{We emphasize that this is not the ADM decomposition of the metric, so in general $\gamma_{ij}$ is not the induced metric on the boundary and $\alpha,\beta_i$ are not the lapse and shift.}
\begin{equation}
\label{canonical}
g= \alpha^2 (dw +\beta_i dx^i)^2 +\gamma_{ij} dx^i dx^j
\end{equation}
where the coordinate $w$ controls the proximity to the boundary or asymptotic region. We may always choose coordinates such that the reference metric is in normal form,
\begin{equation}
\label{canonicalref}
\bar{g}= \bar{\alpha}^2 dw^2 + \bar{\gamma}_{ij} dx^i dx^j \, .
\end{equation}
Now working in the non-coordinate basis, 
\begin{equation}
e^w = dw+\beta_i dx^i \;, \quad e^i =dx^i  \; , \quad e_w = \partial_w  \; , \quad e_i = \partial_i -\beta_i \partial_w  \; ,
\end{equation}
we can calculate,
\begin{eqnarray}
\label{boundaryxiw}
\xi_w &=& \frac{1}{2\alpha^2} \partial_w( \alpha^2 -\bar{\alpha}^2 ) -\frac{1}{2} \gamma^{ij} \partial_w ( \gamma_{ij} -\bar{\gamma}_{ij} )   \\
\xi_i &=& \partial_w \beta_i -\frac{e_i(\alpha^2)}{2\alpha^2} + \frac{\partial_i \bar{\alpha}^2}{2\alpha^2} + \gamma^{kl} ( \Gamma_{ikl} -\bar{\Gamma}_{ikl})   \label{boundaryxii}
\end{eqnarray}
and
\begin{equation}
\label{normsq}
\phi = | \xi |^2 = \frac{1}{\alpha^2} (\xi_w)^2 + \gamma^{ij} \xi_i \xi_j \, .
\end{equation}
In order to use our maximum principle we need $\phi$ to vanish on the boundary, and for regular boundary metric this implies we will need $\xi_w$ and $\xi_i$ to  vanish separately.  

\subsubsection{Manifolds with a boundary}
\label{sec:boundary}

Consider  the case where the manifold has a boundary $\partial \mathcal{M}$. 
For the DeTurck choice of $\xi$ the soliton equation is elliptic, and hence boundary conditions for all components of the metric $g$ should be specified on $\partial\mathcal{M}$.

However, geometrically one considers only imposing boundary conditions on either the induced metric or the extrinsic curvature. In $D$ dimensions the full metric has $D(D+1)/2$ components, and the induced metric or extrinsic curvature each have only $D(D-1)/2$, the difference corresponding to the diffeomorphism freedom normal to the boundary. Thinking of an ADM decomposition with the normal direction to the boundary being `time', the lapse and shift are these non-geometric data which must be fixed in order to give enough boundary conditions for the elliptic Einstein-DeTurck equation. 
These remaining $D$ components are fixed by imposing boundary conditions on the $\xi$ vector. We wish to find Einstein metrics and hence it is natural to require conditions on $\xi$  at the boundary. For example, demanding $\xi$ vanishes provides precisely the additional $D$ boundary conditions needed to specify the lapse and shift.

In the context of the Einstein-DeTurck equation we will now consider two natural sets of boundary conditions that might loosely be termed `modified Dirichlet' and `mixed Dirichlet-Neumann' conditions.

\paragraph{`Modified Dirichlet':}
The most obvious boundary condition would be to fix the induced metric, and also demand that $\xi = 0$. This locally provides the correct number of conditions for the metric components.
However Anderson has shown \cite{Anderson1,Anderson2} the surprising result that fixing these `Dirichlet' boundary conditions does not result in a regular elliptic problem for the Einstein equations modified by a Bianchi gauge fixing term, which is closely related to the DeTurck term we consider here when then metric is near to the reference metric.\footnote{We are grateful to Michael Anderson for pointing out an error in this subsection of a previous version of this work, and for explaining his work  \cite{Anderson1,Anderson2} to us.} In fact more generally Anderson has argued that fixing the induced metric does not result in an elliptic problem for Einstein metrics, and an obstruction to ellipticity is given by the Hamiltonian constraint normal to the boundary. 

Anderson proposes that rather than fixing the entire induced metric, instead one can fix the conformal class of the induced metric, and the remaining freedom is fixed by specifying the trace of the extrinsic curvature. We term these `modified Dirichlet' conditions.
For the Bianchi gauge Einstein equations he has shown that together with requiring $\xi = 0$ this gives a regular elliptic system. Following Anderson's analysis in  \cite{Anderson1} it is a simple exercise to confirm that the Einstein-DeTurck equation is also a regular elliptic system for these boundary data. 

We have therefore suitable elliptic boundary conditions for all metric components of $g$ which precisely fix the conformal class of the induced boundary metric, the trace of the extrinsic curvature and, most importantly from our point of view, ensures that $\phi = 0$ on the boundary. If the only boundary of $\mathcal{M}$ is of this modified Dirichlet type, our maximum principle then implies there are no non-trivial solitons. If there are other types of boundary or asymptotic region, then these modified Dirichlet conditions ensure that if there is a non-trivial soliton, the maximum of $\phi$ must reside on one of these other boundaries or in these asymptotic regions. 
\\

\paragraph{`Mixed Dirichlet-Neumann':} 
We will now consider boundary conditions where we require the induced metric $h_{ij}$ on the boundary to be proportional to the extrinsic curvature $K_{ij}$, as,
\begin{eqnarray}
\label{eq:neumann}
K_{ij} + \lambda \, h_{ij} = 0 \, .
\end{eqnarray}
Such boundary conditions arise in physical problems in the context of codimension one orbifold planes where the tension is related to $\lambda$. An example of such boundaries arise in the Randall-Sundrum braneworld models \cite{RSI,RSII} in the absense of matter on the branes. These orbifold branes may have both positive and negative $\lambda$, corresponding to positive and negative tension.

Let us first consider the case $K_{ij} = 0$ on the boundary. Whilst the manifold really ends at the boundary, in this case we may consider `doubling' the manifold and smoothly gluing it across the boundary. This `doubled' manifold no longer has a boundary but instead a reflection plane so that the metric has a discrete $\mathbb Z_2$ symmetry. This symmetry implies that the normal component of $\xi$ vanishes, and the remaining tangential components of $\xi$ have vanishing normal gradient. Then, provided the reference metric shares the same discrete symmetry, the Einstein-DeTurck tensor on this doubled manifold will also haver this symmetry. With no boundary, the Einstein-DeTurck equation is then clearly elliptic. Furthermore applying our maximum principle to this `doubled' manifold we deduce that $\phi$ could not have a maximum on this reflection plane. Using the metric form \eqref{canonical} where $w = 0$ is the reflection plane, we see that $\alpha$ and $\gamma_{ij}$ will be even functions of $w$, and $\beta_i$ odd. Thus $\beta_i$ vanishes at $w=0$, and hence then the induced metric $h_{ij} = \gamma_{ij}$ on the boundary. Thus $K_{ij}=0$ becomes $\partial_w \gamma_{ij}=0$ at $w=0$.
Likewise for the reference metric we require $\bar{\alpha}$ and $\bar{\gamma}_{ij}$ are even. We then explicitly see from \eqref{boundaryxiw} that $\xi_w = 0$ at $w=0$ and that $\xi_i$ is even in $w$.
This example, $K_{ij} = 0$ at $w=0$, is instructive as it shows that in this case, instead of imposing $\xi_{\mu} = 0$ on the boundary $w = 0$, it is more natural to fix $\xi_w = 0$ and $\beta_i=0$ at $w=0$ (which is consistent with the reflection symmetry of the doubled manifold introduced above).

We now consider boundary conditions for our more general mixed Neumann-Dirichlet problem \eqref{eq:neumann}. We take our metric to be of the form  \eqref{canonical} with the chart covering the manifold for $w \ge 0$ and the boundary being $w = 0$, so the outer unit normal $\partial_n=-\frac{1}{\alpha} \partial_w$.  We begin by requiring that our reference metric \eqref{canonicalref} also obeys the same condition \eqref{eq:neumann} as the metric, which implies $\bar{K}_{ij} = - \frac{1}{2 \bar{\alpha} } \partial_w \bar{\gamma}_{ij} = -\lambda \, \bar{\gamma}_{ij}$.
We then impose the following boundary conditions on the metric \eqref{canonical} at $w = 0$;
\begin{eqnarray}
\label{eq:neumannbc}
\xi_w & = & \frac{1}{\alpha^2} \left( \alpha \partial_w \alpha - \bar{\alpha} \partial_w \bar{\alpha} \right) - \lambda \left( \alpha (m-1) - \bar{\alpha} \gamma^{ij} \bar{ \gamma}_{ij} \right) = 0 \nonumber \\
 \beta_i & = & 0 \nonumber \\
K_{ij} + \lambda  \, h_{ij} & = & - \frac{1}{2 \alpha} \partial_w \gamma_{ij} + \lambda  \, \gamma_{ij} = 0
\end{eqnarray}
where $m$ is the dimension of the manifold. We emphasize that since $\beta_i = 0$ at $w=0$, then the induced metric $h_{ij}=\gamma_{ij}$ and the extrinsic curvature $K_{ij}= -\frac{1}{2\alpha} \partial_w \gamma_{ij}$ at $w=0$. Notice there are no tangential derivatives in these boundary conditions due to the condition $\beta_i=0$. In order to establish elliptic regularity of the system\footnote{
For a reference discussing regularity of PDEs see the textbook of Taylor \cite{Taylor}.
}
it is sufficient to consider the leading order derivative terms in the Einstein DeTurck and boundary operators acting on $\alpha$, $\beta_i$ and $\gamma_{ij}$ (the former of course acts diagonally). In other words, one must consider the short distance behaviour of the field equations and boundary conditions. Provided $\alpha \ne 0$ (as required for a smooth boundary geometry), this yields,
\begin{eqnarray}
\partial_w \alpha \simeq 0 \; , \quad \beta_i \simeq 0 \; , \quad \partial_w \gamma_{ij} \simeq 0 
\end{eqnarray}
where by $\simeq 0$ we mean the righthand side only contains terms without derivatives (which may be ignored at short distances). Then we see that linearised perturbations of this system satisfy Dirichlet conditions for the components of $\beta_i$ and Neumann conditions for $\alpha$ and the components of $\gamma_{ij}$.  Therefore these boundary conditions give a regular elliptic boundary value problem.

A straightforward calculation shows that at the boundary $w = 0$ the tangential components of the vector $\xi$ obey the simple condition,
\begin{eqnarray}
\label{eq:neumanxitang}
\partial_n \xi_i  =  - 2 \lambda \,  \xi_i 
\end{eqnarray}
 where to derive this expression one must use the boundary conditions above, together with the Einstein-DeTurck equations. Together with $\xi_w = 0$ these imply that,
\begin{eqnarray}
\partial_n \phi = - 2 \lambda \phi \, .
\end{eqnarray}
In the case that $\lambda = 0$ or $\lambda > 0$, the zero or positive tension cases for Randall-Sundrum branes, then our maximum principle dictates that if a non-trivial soliton solution exists, the maximum of $\phi$ cannot reside on these boundaries, since the boundary conditions imply the outer normal gradient is zero or negative, whereas the maximum principle states at a maximum on a boundary the outer normal gradient must be strictly positive. If there are no other boundaries or asymptotic regions in the problem then this implies no non-trivial solitons can exist and hence any solution must be Einstein. If there are other boundaries or asymptotic regions this implies that if a non-trivial soliton exists, the maximum of $\phi$ must reside on one of these other boundaries or in the asymptotic regions.

We note that for the case of a negative tension orbifold plane, so that $\lambda < 0$, then the boundary condition does not rule out the existence of a maximum of $\phi$ on the boundary. Obviously \eqref{eq:neumanxitang} is compatible with finding Einstein metrics where $\xi = 0$ everywhere, but our maximum principle can not be used to rule out solitons in this case. In a numerical setting if a solution is found, one must explicitly check whether it is a soliton or not by computing $\phi$.

We note that Anderson has shown \cite{Anderson1} that in the closely related Einstein-Bianchi equation the Neumann boundary condition together with $\xi = 0$ is not regular. One can check this is the case also for the Einstein-DeTurck equation. We emphasize that for our boundary condition above we do not impose $\xi = 0$, but only the normal component $\xi_w = 0$, and this is the reason we have regularity. The elegant behaviour \eqref{eq:neumanxitang} for the tangential components of $\xi$, which is consistent with finding Einstein solutions, requires our specific form of boundary condition, and presumably does not generalize to cases such as $K_{ij} = s_{ij}$ for some tensor $s_{ij}$ on the boundary, unless $s_{ij}$ obeys special properties. Hence one cannot use our mixed Neumann-Dirichlet condition above to study the general Neumann problem with sources.

\subsubsection{Asymptotically Euclidean and Kaluza-Klein manifolds}
\label{sec:AKK}

We now consider asymptotic behaviour of Riemannian manifolds that is appropriate to describe the Euclidean continuation of static asymptotically flat or Kaluza-Klein Lorentzian spacetimes, with or without a non-extremal horizon (so that Euclidean time may or may not be periodic). 

Consider the case where $\Lambda = 0$ and $(\mathcal{M},g)$ is a non-compact asymptotically Euclidean (AE)  or asymptotically Kaluza-Klein (AKK) Riemannian manifold (where for the sake of generality we allow for a general compact Ricci flat internal manifold).  
By definition (see e.g.~\cite{DaiKK}), such a manifold $\mathcal{M}$ has an end diffeomorphic to $E \cong (\mathbb{R}^{m-k} - B_R) \times X_k$,  where $B_R$ is a ball of radius $R$ and $X_k$ is a compact $k$-manifold which admits a Ricci flat metric $g_X$. 
For the AE case,  $k = 0$ and the end is simply diffeomorphic to $E \cong (\mathbb{R}^{m} - B_R)$.
Furthermore, if we let  $r=\sqrt{\delta_{ij} x^{i} x^{j}}$ where $x^i$ are standard Cartesian coordinates on $\mathbb{R}^{m-k}$, the metric $g$ in $E$, i.e. for $r>R$, satisfies
\begin{equation}
g=\delta + g_X  +O(r^{-p})  \; ,   \qquad  \nabla^0 g=O(r^{-p-1})  \; , \qquad    \nabla^0 \nabla^0 g= O(r^{-p-2})
\end{equation}
where $\delta=\delta_{ij} dx^i dx^j$ is the Euclidean metric on $\mathbb{R}^{m-k}$,  $\nabla^0$ is the metric connection of $g_0=\delta+g_X$,  and $p>0$.  
The AE case, the special case of AKK with $k=0$, is of relevance as it gives the correct asymptotics for static Lorentzian spacetimes with no non-extremal horizon, continued to Euclidean signature so that Euclidean time is non-compact.
The AKK case with $k=1$ and $X_k=S^1$ is of physical relevance since it gives the correct asymptotics for a static non-extremal black hole continued to Euclidean time 
where $X_k$ is the Euclidean time circle. The AKK case with $k > 1$ may describe the Euclidean continuation of static Lorentzian spacetimes with Kaluza-Klein asymptotics, possibly with a non-extremal horizon in which case $X_k$ contains an $S^1$ factor.
We will now treat all these cases $k \geq 0$ simultaneously.

The inverse metric satisfies $g^{-1}=g_0^{-1}+O(r^{-p})$. Now introduce coordinates $x^\mu$ for $g$ by $x^\mu=(x^i, y^a)$ where $y^a$ are coordinates on $X_k$.   It easily follows that the Christoffel symbols $\Gamma^\mu_{\nu \rho}=O(r^{-p-1})$ if at least one of the coordinate indices belong to the Cartesian ones $x^i$, and that $\Gamma^a_{bc}= \Gamma(g_X)^a_{bc} + O(r^{-p-1})$. Thus assuming our reference metric $\bar{g}$ is also AKK with the same $g_X$, we easily see
\begin{equation}
\xi^{\mu}= O(r^{-p-1}) \,,\qquad \phi= O(r^{-2p-2})\,,
\end{equation}
and thus $\phi \to 0$ at asymptotic infinity.  We note that in fact we did not need the precise value of $p$ and the argument works for any $p>0$.

If there are no other boundaries/asymptotic regions, a simple argument now allows us to use the maximum principle to rule out non-trivial solitons.
Take $\mathcal{M}$ minus the end $E \cong (\mathbb{R}^{m-k} - B_r)\times X_k$ where $r \geq R$. Denote this manifold $Y_r$, and note that it is compact with $(m-1)$-sphere boundary $\partial Y_r = S^{m-1}$. The AKK asymptotics imply $\phi|_{\partial Y_r} \leq
Cr^{-2p-2}$ for some positive constant $C$. The maximum principle applied to $Y_r$ thus implies $\phi \leq
Cr^{-2p-2}$ everywhere within $Y_r$.  Now, suppose there is a non-zero value of
$\phi$, say $\phi_0>0$, at some point $q$ in $\mathcal{M}$. By taking a large enough $r$
we can always arrange $q$ to be in the interior of $Y_r$ \emph{and} for $Cr^{-2p-2}
<\phi_0$. We therefore have a contradiction and thus we must have $\phi=0$ everywhere.

If there are other boundaries/asymptotic regions, this argument implies that if a non-trivial soliton exists on $\mathcal{M}$, then the maximum of $\phi$ must reside on/in a different boundary/asymptotic region.

For completeness we record how to write AE metrics in the coordinate system (\ref{canonical}).  First write the Euclidean metric in polar coordinates $\delta= dr^2 + r^2 d\Omega^2$.  Then define $\rho =1/r$ so that $\rho \to 0$ corresponds to $r \to \infty$. 
The metric (and similarly the refererence metric) can then be written in the form
\begin{equation}
\begin{aligned}
g &= \frac{\alpha^2}{\rho^4} \left( d \rho +  \omega_i dx^i \right)^2 + \frac{1}{\rho^2} h_{ij} dx^i dx^j \,,
\end{aligned}
\end{equation}
together with requiring the `Dirichlet' conditions, $\alpha = 1$, $\omega_i = 0$ and $h_{ij} =  \Omega_{ij}$ at $\rho = 0$. In fact, more precisely from the above we deduce $\alpha=1+O(\rho^p)$, $\omega_i = O(\rho^{p+2})$ and $h_{ij}=\Omega_{ij} + O(\rho^{p+2})$. 
This is simply generalized to the AKK case.

\subsubsection{Asymptotically locally hyperbolic manifolds}
\label{sec:ALH}

Next we consider the case where $\Lambda=-\frac{m-1}{\ell^2} < 0$ and $(\mathcal{M},g)$ is a non-compact  asymptotically locally hyperbolic (ALH) manifold. We start with the standard conformal definition (see \cite{Ashtekar:1984zz}) which requires the existence of:  i) a Riemannian manifold with a boundary $(\hat{M},\hat{g})$ such that $\mathcal{M}$ is diffeomorphic to $\hat{{M}}-\partial \hat{{M}}$;  ii) a defining function $z$ on $\hat{M}$, such that $z>0$ and $\hat{g}=z^2 g$ in the interior, and $z=0$ and $dz \neq 0$ on $\partial \hat{M}$.  For Einstein manifolds this is then of course supplemented by $\text{Ric}(g)=\Lambda g$ in $\mathcal{M}$. Instead we will consider the Einstein-DeTurck soliton equation (\ref{eq:soliton}) such that the DeTurck vector $\xi$ is constructed from a reference metric $\bar{g}$ in the same class.

First define a vector field $Z$ in $\hat{M}$, by the properties that on $\partial \hat{M}$ it is normal (with respect to $\hat{g}$) and $\hat{g}(Z,Z)=\ell^2$ on $\partial \hat{M}$, such that $Z^\mu \partial_\mu z=1$ in $\hat{M}$. Let $X$ be any vector field tangent to hypersurfaces of constant $z$ (which includes $\partial \hat{M}$), so one has $X^\mu \partial_\mu z=0$ and $[X,Z]=0$. Since $dz \neq 0$ on $\partial \hat{M}$ we may use $z$ as a coordinate. We now define a notion of ALH, sufficiently general for the Einstein-DeTurck equation, by requiring the metric $\hat{g}$ satisfies
\begin{equation}
\label{hatg}
\hat{g} = \ell^2( dz^2 +h^0)  +O(z^p)
\end{equation}
where $h^0$ is the boundary metric (an $(m-1)$-dimensional Riemannian metric), for some $p>0$, together with derivative conditions
\begin{equation}
\label{derhatg}
( \nabla^0_Z )^{n_1} (\nabla^0_X)^{n_2} \hat{g}=O(z^{p-n_1})
 \end{equation}
where $\nabla^0$ is the metric connection of $g^0=\ell^2(dz^2 +h^0)$ and $n_i$ are non-negative integers such that $n_1+n_2 \leq 2$.  These conditions imply that the Riemann tensor of $g= z^{-2} \hat{g}$ approaches that of the standard hyperbolic space $\mathbb{H}^m$ as $ z \to 0$.  More precisely, one can show that\footnote{If we further assume that $p\geq 2$, as for $C^2$-functions near the boundary, then one has $| \, \text{Ric}(g) -\Lambda g \, |^2= O(z^{4})$ and $| \, \text{Weyl}(g) \, |=O(z^{4})$, independently of $p$.}
\begin{equation}
| \, \text{Ric}(g) -\Lambda g \,  |^2= O(z^{2p}) \qquad \qquad | \, \text{Weyl}(g) \, |^2=O(z^{2p})  
\end{equation}
where the norms are of course taken with respect to the metric $g$.
It is thus in this sense the metric is asymptotically hyperbolic.  Observe that from the Ricci soliton equation it immediately follows that $| \nabla \xi |^2 =O(z^{2p})$. Shortly we will show that in fact the stronger statement $|\xi|^2 =O(z^{2p})$ holds.

Therefore, equivalently we may define our notion of an ALH manifold as having an end diffeomorphic  to a quotient of $\mathbb{H}^m - \{ z \geq z_0 \}  $, such that $g$ can be written for $0<z < z_0$ as $g=z^{-2} \hat{g}$ with $\hat{g}$ given by (\ref{hatg}) and (\ref{derhatg}). We will also choose the reference metric $\bar{g}$ to be in the same class with the same boundary metric.

For practical purposes it is convenient to introduce an explicit general chart $(z,x^i)$ valid near the boundary, so $Z=\partial_z$ and $X=\partial_i$, which allows one to write $g$ for small $z>0$ as
\begin{equation}
g= \frac{\alpha^2}{z^2} \left[ \left( dz + \omega_i dx^i \right)^2 + h_{ij} dx^i dx^j \right]
\end{equation}
where $\alpha, \omega_i, h_{ij}$ are all functions of $(z,x^i)$, $z \to 0$ is the asymptotic region, with $\omega_i \to 0$ and $h_{ij} \to h^0_{ij}$. Given (\ref{derhatg}) it follows that the $z\to 0$ behaviour of these functions is
\begin{equation}
\begin{aligned}
& (\partial_z)^{n_1} (\partial_i)^{n_2} ( \alpha-\ell) = O(z^{p-n_1})\,, \qquad (\partial_z)^{n_1} (\partial_i)^{n_2}  \omega_i = O(z^{p-n_1}) \;, \\
& (\partial_z)^{n_1} (\partial_k)^{n_2}  (h_{ij}-h^0_{ij}) =  O(z^{p-n_1}) 
\end{aligned}
\end{equation}
where again $n_i$ are non-negative integers such that $n_1+n_2 \leq 2$. 
Note that in such coordinates we cannot assume smoothness (of $\hat{g}$) near $z=0$, since this property is not even satisfied by ALH Einstein metrics in the Fefferman-Graham expansion (they are differentiable to some finite order determined by $m$ though). 

We remark that from the point of view of an elliptic boundary value problem we may regard taking this behaviour for $\alpha$, $\omega_i$ and $h$ as fixing Dirichlet data for all components of the metric $g$ at the conformal boundary, i.e. $\alpha=\ell$, $\omega_i=0$ and $h_{ij}=h_{ij}^0$ at $z=0$.

As stated above we take the reference metric $\bar{g}$ also to be ALH in the above sense, with the same boundary metric $h^0$ in the conformal frame defined by $z$. Furthermore, we use the coordinate freedom in the chart $(z,x^i)$ and use Gaussian normal coordinates so $\bar{\alpha}=\ell$ and $\bar{\beta}_i=0$. We thus just have
\begin{equation}
\bar{h}_{ij} = h^0_{ij} + O(z^p)  \; .
\end{equation}

The metric and reference metric may be recast to the form \eqref{canonical} by changing to the proper coord $w= - \log z$ and setting $\beta_i = \omega_i / z$, $\gamma_{ij} = h_{ij} / z^2$ and $\bar{\gamma}_{ij} = \bar{h}_{ij} / z^2$.
With our choice of asymptotic form for $g$ and $\bar{g}$, it is then straightforward to check from \eqref{boundaryxiw}, \eqref{boundaryxii} that as $z \to 0$
\begin{eqnarray}
\xi_w = O(z^p) \; , \quad \gamma^{ij} \xi_i \xi_j = z^2 h^{ij} \xi_i \xi_j = O(z^{2p}) \; , 
\end{eqnarray}
and therefore from \eqref{normsq} that
\begin{equation}
\phi  = O(z^{2p})   \; .
\end{equation}
It follows that  $\phi  \to 0$ as $z \to 0$. 

Now, assuming there are no other asymptotic regions of the manifold, and that there are no other boundaries, we can use an analogous argument to the one for the AE/AKK case above to rule out the existence of non-trivial Einstein-DeTurck solitons. Namely, consider the compact manifold with boundary given by $X_\epsilon=\mathcal{M}-E_\epsilon $ where $E_\epsilon$ is the end corresponding to $0<z \leq \epsilon$. The ALH conditions imply $\phi|_{\partial X_\epsilon} \leq C \epsilon^p$ for some positive constant $C$. The maximum principle applied to $X_\epsilon$ thus implies $\phi \leq C \epsilon^p$ everywhere in $X_\epsilon$. Now, suppose there is a non-zero value of $\phi$ at some point $q$ in $\mathcal{M}$. By taking small enough $\epsilon$ we can always arrange $q$ to be in the interior of $X_\epsilon$ and for $C \epsilon^p<\phi(q)$. This gives a contradiction and we thus must have $\phi=0$ everywhere. If there are other boundaries, these asymptotics imply a non-trivial Ricci soliton must have the maximum of $\phi$ on another boundary or asymptotic region.
 
\subsubsection{Asymptotically extremal manifolds}
\label{sec:extremal}

We are actually interested in static {\it Lorentzian} metrics. In this context, we are in particular interested in solutions with Killing horizons, such as black holes. In the Euclidean section, a non-extremal horizon simply appears as a smooth degeneration of the static Killing vector field $\partial / \partial \tau$ (see \S\ref{sec:horizon}), which is neither a boundary or an asymptotic end of the manifold. However, extremal horizons are qualitatively different, and from an Euclidean point of view should be thought of as asymptotic regions of $\mathcal{M}$. More precisely, we will introduce the notion of asymptotically extremal manifolds, in the context non-compact, static Riemannian manifolds. These will be defined in such a way that their Lorentzian sections describe the exterior to a general static spacetime containing an extremal Killing horizon, as we now discuss.

Consider a spacetime containing a smooth static extremal Killing horizon, of a Killing vector field $V=\partial /\partial v$. We will assume that the cross-sections of the horizon $\mathcal{H}$ are simply connected and that $V$ is timelike just outside the horizon (as is the case for static black holes). It has been shown that the {\it near-horizon geometry} associated to such spacetimes can be written as~\cite{KLR}
\begin{equation}
\label{NHG}
g_{NH}= T_0(x) \left[ - \varrho^2dt^2 +\frac{d\varrho^2}{\varrho^2} \right] +\gamma^0_{ab}(x)dx^a dx^b
\end{equation}
where $\varrho=0$ is the horizon, $\gamma^0_{ab}$ is the induced metric on $\mathcal{H}$ and $T_0(x)$ is a positive function on $\mathcal{H}$. The 2d metric in the square brackets is of course AdS$_2$, and so the near-horizon geometry is simply a {\it warped} product of AdS$_2$ and a Riemannian metric $\gamma^0_{ab}$ on $\mathcal{H}$. Although the above metric is strictly only valid for $\varrho>0$,  $\varrho=0$ is merely a coordinate singularity which may be removed by changing coordinates to $(v,\varrho)$ defined by $t=v+\varrho^{-1}$.\footnote{Standard Gaussian null coordinates are given by $(v,r,x^a$) where $r=T_0(x) \varrho$, see Appendix \ref{sec:AppExtremal}.} 

We are of course interested in the full spacetime metric near the extremal horizon, not just its near-horizon limit. 
It can be shown that outside the horizon we may write the spacetime metric $g$ as
 \begin{equation}
 \label{extremalhorizon}
g= -T(\varrho,x) \varrho^2 dt^2+R(\varrho,x) \left( \frac{d\varrho}{\varrho} + \varrho \, \omega_a(\varrho,x) dx^a \right)^2   +\gamma_{ab}(\varrho,x) dx^a dx^b 
\end{equation}
for $\varrho>0$, where $T,R>0$ and $T, R, \omega_a, \gamma_{ab}$ are {\it smooth} functions of $(\varrho,x)$ at $\varrho=0$ (at least),  such that
\begin{eqnarray}  
&&T|_{\varrho=0}=R|_{\varrho=0} =T_0(x) \,,\qquad  \qquad \gamma_{ab}|_{\varrho=0} =\gamma^0_{ab}(x) \label{NH} \\ 
\label{smoothness}
&& \psi(x) \equiv \frac{T_1(x)-R_1(x)}{T_0(x)}  = \text{constant}
\end{eqnarray} 
where $T_1=\partial_{\varrho}T|_{\varrho=0}$  and $R_1=\partial_{\varrho}R|_{\varrho=0}$.
In Appendix \ref{sec:AppExtremal} we prove this starting from a general smooth extremal Killing horizon written in Gaussian null coordinates, and then imposing staticity.  

It is worth noting that that {\it given} the metric (\ref{extremalhorizon}), which appears singular at $\varrho=0$, it is easy to see the conditions (\ref{NH}) and (\ref{smoothness}) are sufficient for the existence of a smooth extremal Killing horizon at $\varrho=0$ whose near-horizon limit coincides with (\ref{NHG}).
Explicitly, consider coordinate transformations of the form
\begin{equation}
\label{coords:extremal1}
dt = dv +\left( \frac{a_0}{\varrho^2} +\frac{a_1}{\varrho} \right) d\varrho
\end{equation}
for some constants $a_0,a_1$ which are chosen to ensure that the metric functions and the inverse metric functions are smooth at $\varrho=0$. One finds that 
\begin{equation}
\label{coords:extremal2}
a_0=-1\;, \qquad  \qquad a_1= \frac{T_1-R_1}{2T_0}   \; .
\end{equation}
However, since $a_0,a_1$ must be constants we learn a non-trivial condition that must be met to ensure smoothness of the extremal horizon, namely (\ref{smoothness}). Assuming this is the case the metric near $\varrho=0$ then looks like
\begin{equation}
g= T_0(x) [ -\varrho^2 (1+O(\varrho) ) dv^2+ (1+O(\varrho)) dv d\varrho ] + 2R  \, \omega_a dx^a d\varrho+ \gamma_{ab}dx^a dx^b 
\end{equation}
which is indeed smooth and invertible at $\varrho=0$. This allows one to extend to the region behind the horizon $\varrho<0$. The surface $\varrho=0$ is then a smooth extremal Killing horizon of $\partial / \partial v$ as claimed.  Finally, the near-horizon geometry defined by setting $\varrho \to \epsilon \varrho $ and $t\to t/ \epsilon$ and taking $\epsilon \to 0$ coincides with (\ref{NHG}) as required.

Next, consider Euclideanising the geometry by setting $t=i\tau$. Now  $\varrho \to 0$ corresponds to a new asymptotic region. This motivates the following definition: we will say that a non-compact, static, Riemannian manifold ($\mathcal{M}, g, \partial /\partial \tau)$ is {\it asymptotically extremal}, if $\mathcal{M}$ has an end in which the metric $g$ can be written in coordinates $(\tau, \varrho, x^a)$ such that
 \begin{equation}
 \label{extremalend}
g= T(\varrho,x) \varrho^2 d\tau^2+R(\varrho,x) \left( \frac{d\varrho}{\varrho} + \varrho \, \omega_a(\varrho,x) dx^a \right)^2   +\gamma_{ab}(\varrho,x) dx^a dx^b 
\end{equation}
for sufficiently small $\varrho>0$, where $T,R>0$ and $T,R, \omega_a, \gamma_{ab}$ are smooth at $\varrho=0$ and the conditions (\ref{NH}) and (\ref{smoothness}) are satisfied.
We may consider either non-compact or periodic Euclidean time. We emphasize that in either case the Riemannian geometry will be smooth in the interior, even though in the periodic case the time circle degenerates as $\varrho \rightarrow 0$, and that by construction the Lorentzian continuation will be smooth at $\varrho = 0$ (and indeed one can continue through the horizon).

Again, the class of metrics (\ref{extremalend}) may be cast in the form \eqref{canonical} by defining the proper coordinate $w=\log \varrho$ and $x^i=(\tau, x^a)$ and setting
 \begin{eqnarray}
 \alpha^2 = R\;,  \qquad \beta_idx^i = \varrho \, \omega_a dx^a  \; , \qquad
 \gamma_{ij}dx^idx^j = T \varrho^2 d\tau^2 +\gamma_{ab} dx^a dx^b  \; .
 \end{eqnarray}
We also assume the reference metric $\bar{g}$ is in the same class as $g$, with the same near-horizon geometry as $g$. That is, as $\varrho \to 0$ the metric $\bar{g}$ can also be written as (\ref{extremalend}) for some $\bar{T}, \bar{R}$, $\bar{\gamma}_{ab}$  and $\bar{\omega}_a=0$ (choosing normal coordinates), such that $\bar{T}_0=\bar{R}_0=T_0$ and $\bar{\gamma}_{ab}=\gamma_{ab}$. For such $g$ and $\bar{g}$, it is again straightforward to show from \eqref{boundaryxiw}, \eqref{boundaryxii} and \eqref{normsq} that as $\varrho \to 0$
\begin{eqnarray}
\xi_w = O(\varrho) \; , \quad \xi_\tau = 0 \; , \quad \xi_a = O(\varrho) \;,\quad  
\phi = O(\varrho^{2})
\end{eqnarray}
and hence $\phi \to 0$ as $\varrho \to 0$. One can then use an analogous argument to the ALH case, to show the maximum principle rules out  the existence of non-trivial solitons, unless there are other asymptotic regions or boundaries where $\phi$ may achieve a maximum.

\subsubsection{Fictitious boundaries}
\label{sec:horizon}

It is convenient to adapt coordinates to any isometries a geometry may have. Since
there will be no dependence of the metric components on the coordinates corresponding to the symmetry directions, 
the effective dimension of the Einstein-DeTurck PDE system is reduced. However, if the isometry does not have a free action, these adapted coordinates will not cover the fixed points. The canonical example of this is polar coordinates which strictly speaking do not cover the origin point. One possibility is to use other charts to cover the neighbourhood of these fixed points that are not adapted to the symmetry. However, a more practical option is to treat these fixed point sets as `fictitious boundaries' for the adapted coordinate chart, and to determine the necessary `boundary conditions' in these adapted coordinates by deducing them from smoothness in a chart that does cover the fixed point set. Since these are not real boundaries of the manifold then of course our maximum principle implies that for a non-trivial soliton solution, the maximum of $\phi$ cannot reside at these fictitious boundaries. For a black hole the horizon and any axes of rotational symmetry are typical examples of submanifolds that are usefully treated as fictitious boundaries for coordinates adapted to static and rotational symmetry. We shall discuss these examples below, giving the fictitious boundary conditions to be imposed in these adapted coordinates, and also demonstrating how these conditions imply that a maximum of $\phi$ cannot occur.

\paragraph{Non-extremal horizons:} First consider a static spacetime containing a generic smooth static non-extremal horizon, such that the static vector is timelike just outside the horizon. Outside the horizon, the metric may be written in coordinates adapted to the static isometry $(t,w,x^a)$ and takes the form
\begin{equation}
\label{nonextremehorizon}
g= - T^2 w^2 dt^2+ W^2 \left( dw + w\,  \Omega_a dx^a \right)^2  + \gamma_{ab} dx^a dx^b 
\end{equation}
for $w>0$, where the horizon is at $w=0$, such that $T, W>0$ and $T, W, \Omega_a, \gamma_{ab}$ are smooth functions of $w^2$  and $x^a$ at $w=0$ and
\begin{equation}
\label{reghor}
\left. \frac{T}{W} \right|_{w=0}  = \kappa
\end{equation}
where $\kappa>0$ is a constant equal to the surface gravity. A proof of this is provided in  Appendix \ref{sec:AppNonExtremal}.  Conversely, given the metric (\ref{nonextremehorizon}), which appears singular at $w=0$, it is easy to see that the smoothness conditions just mentioned together with (\ref{reghor}), are sufficient for the existence of a smooth non-extremal Killing horizon at $w=0$ with surface gravity $\kappa$. To see this explicitly, consider the coordinate transformation
\begin{equation}
r=w^2\;, \qquad  v = t +\frac{1}{\kappa} \log w
\end{equation}
in which case the metric near $r=0$ is
\begin{equation}
g=\frac{T^2}{\kappa} \left[ -r \kappa dv^2+ dvdr \right] +O(1) dr^2 + W^2 \Omega_a dr dx^a  +(\gamma_{ab} +W^2 r \Omega_a \Omega_b) dx^a dx^b
\end{equation}
with all functions smooth at $r=0$. This shows that the metric at $r=0$ is smooth and invertible and allows one to extend to the region $r<0$. Therefore as claimed the surface $r=0$ is a smooth non-extremal Killing horizon of $\partial / \partial v$ with surface gravity $\kappa.$

Now, consider Euclideanising this geometry by setting $t=i\tau$, so
\begin{equation}
\label{staticaxis}
g= T^2 w^2 d\tau^2+ W^2 \left( dw + w\,  \Omega_a dx^a \right)^2  + \gamma_{ab} dx^a dx^b 
\end{equation}
for $w>0$ such that $T, W>0$ and $T, W, \Omega_a, \gamma_{ab}$ are smooth functions of $w^2$ and $x^a$ at $w=0$ and (\ref{reghor}) holds. 
The static Killing field $\partial / \partial \tau$ now vanishes at $w=0$ where the isometry it generates has fixed action.
However, due to the condition (\ref{reghor}) this Riemannian metric is guaranteed to be smooth at $w=0$, provided one identifies $\tau$ periodically
\begin{equation}
\label{tauperiod}
\tau \sim \tau + \frac{2\pi}{\kappa}   \; .
\end{equation}
This is easy to see by changing from the 2d  ``polar" coordinates $(w,\tau)$, to Cartesian ones $(x,y)$ defined by 
$x= w \cos \kappa \tau$ and $y=w \sin \kappa \tau$,
in the usual way. Therefore from the Euclidean point of view, a spacetime containing a  static smooth non-extremal horizon corresponds to a smooth Riemannian manifold with a static $U(1)$ isometry with a fixed action on a  co-dimension-2  submanifold.  Conversely, from the above it is easy to show that any smooth Riemannian manifold with a static $U(1)$ isometry with a fixed action on a  co-dimension-2  submanifold, has a Lorenzian continuation which  can be extended to a static spacetime with a smooth non-extremal horizon.

This metric may be cast in the form \eqref{canonical} by taking $x^i=(\tau, x^a)$ and setting 
 \begin{eqnarray}
\alpha= W ,\qquad \beta_a = w \Omega_a  \; , \quad \gamma_{ij}dx^idx^j = T^2 w^2 d\tau^2 +\gamma_{ab} dx^a dx^b  \; .
 \end{eqnarray}
We take the reference metric $\bar{g}$ to be in the same class, so,
  \begin{eqnarray}
  \bar{\alpha} =\bar{W}, \qquad \bar{\gamma}_{ij}dx^idx^j &=& \bar{T}^2 w^2 d\tau^2 + \bar{\gamma}_{ab} dx^a dx^b  \; .
 \end{eqnarray}
 with the same condition that $\bar{T}, \bar{W}, \bar{\gamma}_{ab}$ are smooth in $x$ and $w^2$ at $w=0$. 
From \eqref{boundaryxiw}, \eqref{boundaryxii} and \eqref{normsq} we see the smoothness of the functions $T, W, \Omega_a, \gamma_{ab}$ in $w^2$, and likewise for the reference metric, implies that  $\partial_w \phi = 0$ at $w = 0$. This shows how the maximum principle implies there can be no maximum of $\phi$ located at the horizon from the perspective of the ``polar" coordinates. \\
 
\paragraph{Axis of rotation:}  We note that although the $U(1)$ above corresponds to Euclidean time, it could equally originate from a spatial rotational isometry. One can obtain an analogous result for a rotational $SO(n)$ action with $n>2$ that has fixed action on some submanifold. Then we may write the metric in ``polar" coordinates as
 \begin{equation}
g= W^2 \left( dw + w \, \Omega_a dx^a \right)^2  + S^2 w^2 d\Omega^2 + \gamma_{ab} dx^a dx^b 
\end{equation}
for $w>0$, where $d\Omega^2$ is the unit $(n-1)$-sphere line element associated to the $SO(n)$ action. Again, smoothness of the geometry implies $W, S, \Omega_a, \gamma_{ab}$ are smooth functions of $x$ and $w^2$ at $w=0$, the only difference being that 
\begin{equation} 
\left. \frac{S}{W}\right|_{w=0} = 1
\end{equation}
since for $n>2$ the $(n-1)$-sphere has curvature and must shrink at a particular rate. Again, taking the same behaviour for the reference metric,  one finds $\partial_w \phi = 0$ at $w = 0$, and hence as we expect a maximum in $\phi$ is also excluded there.

\section{Ricci-DeTurck flow on manifolds with boundaries}
\label{sec:parabolic}

In this section we discuss Ricci-DeTurck flow as a method for solving the Einstein-DeTurck equation.

Consider a one-parameter family of Riemannian metrics $g$ labelled by $\lambda$, on $\mathcal{M}$. The Ricci-DeTurck flow (with cosmological term) is a diffusive flow of the soliton equation (\ref{eq:soliton}), namely
\begin{eqnarray}
\label{eq:flow}
\frac{\partial g_{\mu\nu}}{\partial \lambda}  & = & -2 \left( R_{\mu\nu} - \nabla_{(\mu} \xi_{\nu)} - \Lambda g_{\mu\nu} \right)
\end{eqnarray}
where $\lambda$ is thought of as an auxiliary flow time. With the DeTurck choice of vector $\xi$ this flow is parabolic, and may be straightforwardly simulated numerically given ``initial data".  A natural choice we will use for this initial data is to take $g|_{\lambda=0}=\bar{g}$, where $\bar{g}$ is the reference metric used to define the DeTurck vector $\xi$ (\ref{eq:xi}), and hence $\xi|_{\lambda =0}=0$.  If there are boundaries/asymptotic regions, then on top of the initial data, one needs to specify boundary/asymptotic conditions to make the problem well-posed, which we will discuss shortly.  We note that it is well known that this flow is diffeomorphic to the Ricci flow $\partial \tilde{g}_{\mu\nu} / \partial \lambda   =  -2 \left( \tilde{R}_{\mu\nu} - \Lambda \tilde{g}_{\mu\nu} \right)$, where the diffeomorphism is generated by the vector field $\xi$. In the case of manifolds with boundaries, the metrics $g$ and $\tilde{g}$ will satify the same boundary conditions if $\xi=0$ on the boundary (since $\xi$ generates the diffeomorphism).  We note that Ricci flows in the presence of certain Dirichlet boundaries have been previously considered in \cite{Headrick:2006ti,Holzegel:2007zz}, and in particular in \cite{Holzegel:2007zz} a proof of uniqueness was given where symmetry reduced the problem to the flow of  a cohomogeneity-one metric. For Neumann boundary conditions existence of Ricci flows has been proven \cite{neumann} when the boundary is totally geodesic.

Recall, that ultimately our motivation is to find new Einstein metrics. As pointed out in the introduction, there are two potential problems with using Ricci-DeTurck flow as a method to find Einstein metrics. The first is simply that if the solution we seek is not a stable fixed point of the flow, then in practice this method is harder to implement. The second is that if we do converge to a fixed point,  then we in fact have a solution to the Einstein-DeTurck soliton equation.  Therefore, if we are to use this flow as an algorithm to generate Einstein metrics, we would wish to avoid or at least identify non-trivial solitons. As we showed in the last section for a variety of boundary and asymptotic conditions on $g$ of interest,  a maximum principle forbids the existence of such Einstein-DeTurck solitons provided we also impose these conditions on the reference metric $\bar{g}$ (and indeed these boundary conditions all possess the property that $\xi=0$ on the boundary).

The key issue in using Ricci-DeTurck flow in the presence of boundaries, is whether the boundary conditions required for the soliton equation are in fact preserved by the flow. This is of course a desirable feature since one can start with a metric which satisfies the boundary conditions of interest, and then if one flows to a fixed point one is guaranteed to have a solution satisfying the required boundary conditions.   We will now discuss this for the types of boundary conditions introduced in \ref{sec:elliptic}.

For general manifolds with a boundary we have discussed ``modified Dirichlet"  and ``mixed Neumann-Dirichlet" type boundary conditions in \S\ref{sec:boundary}. In the modified Dirichlet case we have $\xi=0$ on the boundary. Since $\xi$ generates infinitessimal diffeomorphisms along the Ricci-DeTurck flow, the vanishing of $\xi$ at the boundary implies the boundary points are fixed during the flow. Thus we expect this type of boundary  condition to be preserved along the flow. In the mixed Neumann-Dirichlet case we again have that the normal component of $\xi$ vanishes, ensuring the boundary points are preserved by the flow. We note that in this case diffeomorphisms generated by $\xi$ are allowed along the flow that move the points within the boundary, since the tangential components of $\xi$ are not restricted to vanish.

Let us now address the case of fictitious boundaries considered in \S\ref{sec:horizon}.  In fact, the case of fictitious boundaries which arise from static non-extremal horizons, has been considered previously~\cite{KitchenHeadrickWiseman}. Indeed, it is easy to see the class of metrics (\ref{staticaxis}) satisfying (\ref{reghor}) are preserved by  Ricci-DeTurck flow, using the fact these manifolds are equivalent to the class of smooth Riemannian manifolds with a static $U(1)$ isometry with a fixed action on a codimension 2 submanifold.  We simply note that since Ricci-DeTurck flow preserves isometries, provided no singularities are met the $U(1)$ action must leave the same codimension 2 submanifold fixed.  An analogous argument works for the fictitious boundary that arises from an axis of symmetry.

Now we consider the case where the manifold has an end in which the metric takes some asymptotic form. In fact the asymptotically Euclidean case has been considered before. In particular, Ricci flow on asymptotically Euclidean manifolds has been shown to preserve this class of manifolds \cite{ToddOlnyik}. Presumably this can be generalised to the Kaluza-Klein case. We merely note here that since our soliton boundary conditions ensure $R_{\mu\nu}= O(r^{-p-2})$ and $\nabla_{(\mu} \xi_{\nu)}= O(r^{-p-2})$ (see \S\ref{sec:AKK}), 
as long as the (Ricci flat) metric $g_X$ on the internal manifold is independent of $\lambda$, the Ricci-DeTurck flow equation is satisfied asymptotically.\footnote{In fact naively this argument suggests that the ADM mass does not change along the flow, so one has to take these arguments with caution, see~\cite{ToddOlnyik}.}  This leaves asymptotically locally hyperbolic manifolds and static asymptotically extremal manifolds.  In the ALH case we will only provide a brief discussion, and note that recently, under assumptions of smoothness (which generally do not hold for even boundary dimension), Ricci flow has been shown to preserve ALH metrics \cite{Bahuaud}.
The asymptotically extremal case has not been previously considered and the majority of the remainder of this section will be devoted to proving the flow preserves the class of asymptotically extremal manifolds.

\subsection{Asymptotically locally hyperbolic manifolds}
\label{parabolic:ALH}
An important question which we now address is whether Ricci/Ricci-DeTurck flow preserves the class of ALH manifolds in question. This is crucial in order to guarantee that given an initial ALH manifold, any fixed point of the flow is also an ALH manifold. Bahuaud has recently proved short time existence of Ricci flow that preserves smoothness \cite{Bahuaud}.\footnote{We thank Eric Woolgar for bringing this reference to our attention.} The metric can be written as
\begin{eqnarray}
g = \frac{\ell^2}{z^2} \left( dz^2 + \left( h_{ij}(z, x) + z \, v_{ij}(\lambda,z,x) \right) dx^i dx^j \right)
\end{eqnarray}
for a flow time $\lambda$, so $h_{ij}$ is fixed and $v_{ij}$ varies such that $v_{ij}|_{\lambda = 0} =0$. Hence $h_{ij}$ gives the conformal boundary metric which is fixed in flow time. Then Bahuaud has shown that provided $z^2 g $ is smooth up to the boundary initially, then the Ricci flow exists for short times, and $z^2 g$ remains smooth up to the boundary.

However, in even boundary dimensions the Fefferman-Graham expansion shows that an Einstein metric need not be smooth.
Nevertheless, it seems reasonable to expect that even in the non-smooth case, provided the boundary metric is fixed,  Ricci/Ricci-DeTurck flow does indeed preserve the class of ALH manifolds as defined above in section \ref{sec:ALH}. 
To be more precise we will define ALH Ricci-DeTurck flow in an analogous manner to the Einstein-DeTurck equation in \S\ref{sec:ALH}. That is supplement properties (i) and (ii) of the standard conformal definition of ALH at the beginning of \S\ref{sec:ALH}, together with the defining function $z$ being {\it independent} of flow time $\lambda$, and of course imposing the Ricci-DeTurck flow (\ref{eq:flow}) with $\Lambda< 0$ on $\mathcal{M}$.  The key question is then whether the remaining boundary conditions given in \S\ref{sec:ALH} are preserved by the flow. Assume the boundary conditions in \S\ref{sec:ALH} are satisfied along the flow, so asymptotically as $z \to 0$ we have $\textrm{Ric}(g) \sim \Lambda g$ and $\nabla \xi \to 0$, and therefore from the flow equation $\partial g / \partial \lambda \to 0$. Therefore we learn that a {\it necessary} condition that the flow is well defined is that the conformal boundary metric $h^0$ is fixed along the flow, i.e. it is independent of $\lambda$.  It would be very interesting to show this is also sufficient in the case that the metric is not smooth.

An interesting question is whether there is some analog of the Fefferman-Graham expansion that is valid along the flow, and possibly some notion of boundary stress tensor.
In the Appendix \ref{sec:AppALH} we argue that a naive application of the Fefferman-Graham expansion can not apply along the flow, by studying the simpler problem of diffusion of a scalar in hyperbolic space. It would be very interesting to investigate this further, and determine whether there are suitable asymptotic expansions and boundary stress tensors.

\subsection{Static asymptotically extremal manifolds}
Now we address whether Ricci-DeTurck flow preserves the static asymptotically extremal manifolds, which we defined in \S\ref{sec:extremal}. This question has two parts. The first is simply whether the flow in the class of metrics (\ref{extremalend}) preserves the conditions (\ref{NH}); if so, the second question is whether the smoothness condition (\ref{smoothness}) is preserved by the flow.  

Let us address the first part of the question, i.e whether the flow preserves the near-horizon geometry. Thus, suppose (\ref{extremalend}) now depends on $\lambda$, such that at $\lambda=0$ the near-horizon geometry coincides with (\ref{NHG}). Explictly, the ``near-horizon" limit is extracted by setting  $\varrho \to \epsilon \varrho$ and $ \tau\to  \tau/ \epsilon$ and $\epsilon \to 0$, so that 
the full metric (\ref{extremalend}) becomes:
\begin{equation}
g_0 =\varrho^2 T_0(x;\lambda) d\tau^2 + \frac{R_0(x;\lambda) d\varrho^2}{\varrho^2} +\gamma^0_{ab}(x;\lambda)dx^a dx^b
\end{equation}
subject to the ``initial" conditions $T_0(x;0)= T_0(x)=\bar{T}_0(x)$, $R_0(x;0)= T_0(x)=\bar{T}_0(x)$ and $\gamma_{ab}(x;0)=\gamma^0_{ab}(x)=\bar{\gamma}^0_{ab}(x)$, i.e.  $g_0|_{\lambda=0}= g_{NH} = \bar{g}_0$. The near-horizon limit of the full flow equation gives
\begin{equation}
\label{NHflow}
\frac{\partial g_{0}}{\partial \lambda} = -2 \left( \text{Ric}(g_0) - \frac{1}{2}\mathcal{L}_{\xi_0} g_0 -\Lambda g_0 \right)
\end{equation}
where $\xi_0$ is the DeTurck vector associated to the metric $g_0$ and background metric $\bar{g}_0$ (which is the near-horizon limit of the full DeTurck vector). This can be thought of as a Ricci-DeTurck flow equation for the near-horizon geometry itself. Now, let us assume that our initial near-horizon geometry solves the Einstein equations $(\text{Ric}(g_0)- \Lambda g_{0})|_{\lambda=0}=0$. Also  note that since the initial near-horizon geometry coincides with the reference one we must have $\xi_0|_{\lambda=0}=0$.  Therefore, it is clear that a solution to the near-horizon geometry flow equation (\ref{NHflow}) for $\lambda>0$ is given by taking $T_0(x;\lambda)=T_0(x)=R_0(x;\lambda)$ and $\gamma^0_{ab}(x;\lambda) =\gamma^0_{ab}(x)$. Now, appealing to uniqueness of Ricci-DeTurck flow,\footnote{This has been shown for certain complete non-compact Riemannian manifolds \cite{noncompact1, noncompact2} and we assume it to be the case here.}
 this must be the only solution with such initial conditions. This shows that if one starts with a near-horizon geometry which solves Einstein's equations, then it will not change under the flow. In other words the flow preserves the near-horizon geometry.  From now on we will assume this to be the case.

The second question which needs to be addressed is whether the smoothness condition (\ref{smoothness}) is preserved along the flow. This is a more non-trivial question which requires expanding the full Ricci-DeTurck flow equation to first order in $\varrho$, to derive flow equations for $T_1, R_1$, and hence $\psi$ defined above. The details of this calculation are given in Appendix \ref{sec:AppExtremal}.  We find a remarkably simple flow equation for $\psi$ on $\mathcal{H}$:
\begin{equation}
\frac{\partial \psi}{\partial \lambda} = \nabla^2 \psi +  \frac{dT_0}{T_0} \cdot  d \psi   
\end{equation} 
where $\nabla$ and $\cdot$ are the Levi-Civita connection and contraction with respect to the metric $\gamma^0_{ab}$.
Recall that smoothness of the extremal horizon requires $\psi$ to be a constant (\ref{smoothness}). If we take an initial condition $\psi|_{\lambda=0}= \psi_0$ where $\psi_0$ is a constant, then well-posedness of the flow equation for $\psi$ shows that $\psi=\psi_0$ for all $\lambda>0$. This is simply because, $\psi=\psi_0$ clearly solves the flow equation with such an initial condition,  and by uniqueness this must be the only such solution.  This proves that Ricci-DeTurck flow preserves the class of static asymptotically extremal Riemannian manifolds, or in Lorenzian language, the class of {\it smooth} extremal static horizons with a {\it given} near-horizon solution to Einstein's equation.

\section{Example: AdS$_5$ gravity dual to strongly coupled CFT$_4$ on a Schwarzschild background in the Unruh or Boulware vacua}
\label{sec:example}

In this section we will demonstrate the use of the Ricci-DeTurck flow to find a $5d$ Einstein metric with $\Lambda < 0$. This will be a static axisymmetric spacetime, with asymptotic AdS boundary metric being $4d$ Schwarzschild, and with metric asymptoting to the AdS Poincare horizon far from the boundary. In the context of AdS/CFT, taking a trivial product with a round $S^5$ then gives the classical bulk metric dual to $\mathcal{N} = 4$ super Yang-Mills on a Schwarzschild background. 
The IR boundary conditions we impose (i.e. that the geometry tends to the AdS Poincare horizon) are consistent with the classical geometry describing the dominant bulk saddle point dual to the the CFT in 
the Unruh or Boulware vacua, where asymptotically the theory is in vacuum.
We note that they are different to those considered in \cite{Hubeny:2009ru}, who conjecture the behaviour of the dual geometry for the Hartle-Hawking vacuum, where one has a finite temperature horizon in the IR at equilibrium with the boundary horizon, and one expects the CFT to asymptotically be at finite temperature.
Here we have chosen to study the classical dual describing the CFT Unruh and Boulware vacua as,
i) it nicely illustrates the implementation of boundary conditions for extremal horizons discussed formally above, and ii)
 the physics of the Unruh and Boulware vacua is remarkably different for this strongly coupled CFT than for free field theory. This latter point has important implications for the related issue of existence of Randall-Sundrum braneworld black holes \cite{Tanaka:2002rb, Emparan:2002px,Emparan:1999wa,RandallTW,GregoryRoss,Hubeny:2009ru,Hubeny:2009rc}.

As discussed above we treat the static solution by considering the Euclidean section, which is a Riemannian manifold with two asymptotic regions (one locally hyperbolic, the other extremal) and smooth in the interior.
We will find that a single coordinate chart suffices to cover this manifold, and in the following sections will discuss and motivate its choice. As we shall see, the problem is static and axisymmetric, and hence of cohomogeniety two. It depends non-trivially on two coordinates, essentially the boundary (CFT) radial coordinate, and the bulk coordinate, and compactifying these the problem is formulated on a rectangular domain. Two boundaries are fictitious, associated to the axis of rotational symmetry and the vanishing of $\partial/\partial\tau$ at the horizon. The remaining two are the asymptotic AdS boundary (corresponding to the UV of the CFT) whose conformal boundary metric we specify as Schwarzschild, 
and the asymptotic extremal horizon (corresponding to the IR of the CFT) we choose to have a near horizon geometry given by Poincare AdS$_5$. Using our discussion above we are able to rule out a maximum of $\phi$ on all of these boundaries, and hence the existence of any non-trivial Ricci solitons. Crucially, this implies that provided the DeTurck flow converges to a fixed point, it must give an Einstein metric relevant for our problem. Finally, we numerically simulate the flow and show that indeed a stable fixed point exists.

\subsection{Static and axisymmetric metrics: coordinates for the problem}

First let us consider AdS$_5$ in Poincare coordinates
\begin{equation}
g_{AdS_5}= \frac{\ell^2}{z^2} \left( dz^2 +dR^2 +R^2 d\Omega_{(2)}^2 - dt^2 \right)
\end{equation}
where $z \geq 0$ and $R \geq 0$ (equality is not strictly allowed in these coordinates, and corresponds to the the conformal boundary and the fixed point set of $SO(3)$ respectively), and $\Lambda = -\frac{4}{\ell^2}$.  Now introduce new coordinates $(r,x)$ by
\begin{equation}
z= \frac{1-x^2}{1-r^2}\,, \qquad R = \frac{x\sqrt{2-x^2}}{1-r^2}
\end{equation}
with  $0 \leq x < 1$ and $0 \leq r <1$ to ensure an invertible transformation. The metric in these coordinates is
\begin{eqnarray}
\label{AdS5poincare}
g_{AdS_5}&=& \frac{\ell^2}{(1-x^2)^2} \left( -f(r)^2 dt^2 +\frac{4\,r^2}{f(r)^2}\,dr^2  + \frac{4}{g(x)}\, dx^2 + x^2 g(x)\, d\Omega_{(2)}^2 \right)  \nonumber
\end{eqnarray}
where for covenience we define  
\begin{equation} 
\label{fgdef}
f(r)=1-r^2\,,  \qquad g(x)=2-x^2  \; .
\end{equation}  In these coordinates the conformal boundary is in the asymptotic region $x \to 1$, and $x=0$ is the fixed point set of the $SO(3)$ symmetry. Moreover, these coordinates are particularly adapted to the Poincare horizon which is at $r=1$ and $x<1$. Indeed, for $x<1$, we can define new coordinates $(v,\varrho)$  by $f(r)=\varrho$ and $t= v+ \frac{1}{\varrho}$ to get
\begin{eqnarray}
\label{ads5NH}
g_{AdS_5} &=& \frac{\ell^2}{(1-x^2)^2}\left[  -\varrho^2 dv^2 +2dv d\varrho  +  \frac{4}{g(x)}\,dx^2 +x^2 g(x)\, d\Omega_{(2)}^2 \right]
\end{eqnarray}
which confirms that $\varrho=0$ is a smooth extremal Killing horizon w.r.t. $\partial / \partial v$, and allows one to extend through the horizon to the region $\varrho<0$. This expresses AdS$_5$ as a warped product of AdS$_2$ and a metric $\gamma_{ab}$ on cross-sections of the Poincare horizon, i.e. as a static near-horizon geometry  with {\it non-compact} cross-sections of the horizon.\footnote{Note that the general form for a static near-horizon geometry was derived in \cite{KLR}, and we have just written Poincare AdS in this general form.}  The geometry induced on the Poincare horizon is thus given by
\begin{equation}
\gamma_{ab}dx^a dx^b = \frac{\ell^2}{(1-x^2)^2} \left( \frac{4 }{g(x)}\,dx^2 +x^2 g(x) \,d\Omega_{(2)}^2 \right)  
\end{equation}
where $0 \leq x <1$, which is easily seen to be locally conformal to the standard round metric on $S^3$ (by changing coordinate to $y=1-x^2$). Hence, topologically one can think of spatial sections of the Poincare horizon as  open hemispheres of $S^3$.
Note that as $x \to 0$ the metric is of course regular and looks like the origin of $\mathbb{R}^3$, whereas $x \to 1$ corresponds to an asymptotic region (which touches the conformal boundary).

We wish to find a static solution which is asymptotically locally AdS, with the metric on the conformal boundary in the same conformal class as $4d$ Schwarzschild, such that it approaches the AdS Poincare horizon far from the conformal boundary.  In the Euclidean section our manifold is thus ALH with conformal boundary metric given by Euclidean Schwarzschild which of course has $U(1)\times SO(3)$ isometry. In fact, it has been shown that regular static ALH Einstein metrics must inherit the isometry of the boundary metric~\cite{AndersonChrusciel}. Therefore our full spacetime metric must have $U(1)\times SO(3)$ symmetry. 

We now write an ansatz for such a static Riemannian metric in the coordinates developed above
\begin{eqnarray}
g_5 &=& \frac{1}{z(r,x)^2} \left( r^2 T d\tau^2+ \frac{x^2 g(x) S}{f(r)^2}  d\Omega_{(2)}^2+ \frac{4\,A}{f(r)^4}  dr^2  + \frac{4\, B}{f(r)^2g(x)}\,dx^2  + \frac{ 2\,r\,x F }{f(r)^3}\, dr dx   \right) \nonumber\\
 && z(r,x)  =  \frac{1-x^2}{ 1 - r^2 }   \label{ansatz}
\end{eqnarray}
where $f,g$ are defined by (\ref{fgdef}),  $X = \{ T, S, A, B, F \}$ are smooth functions only depending on the coordinates $(r, x)$, thus making the static isometry generated by $\partial / \partial \tau$ and the spherical symmetry both manifest. As above, the coordinate ranges are $0 \leq x < 1$ and $0\leq r <1$. Eventually we will treat the asymptotic regions $x=1$ (conformal infinity) and $r=1$ (Poincare horizon) as boundaries and work on the square $0 \leq x \leq 1$, $0 \leq r \leq 1$ with $X$ being bounded on this domain (and smooth everywhere except at $x=1$). The other domain boundaries $r = 0$ and $x = 0$ we will consider as fictitious ones associated to the (non-extremal) black hole horizon and the axis of rotational symmetry.\footnote{We note that after choosing the reference metric there is in principle sufficient coordinate freedom in the Einstein solutions of the Einstein-DeTurck equation to choose these boundaries of the coordinates \cite{KitchenHeadrickWiseman}. This is reflected in the Ricci-DeTurck flow by the vanishing of the normal component of $\xi$ at the edge of this domain which follows from the various boundary conditions.}
Of course, we require that $T,S>0$ and $AB> r^2 x^2 g(x)F^2/16$ (i.e. the determinant of the $r,x$ part of the metric is positive), to ensure we have a good Riemannian metric. We note that for $r^2 T = S=A=B=1$ and $F=0$ the metric reduces to AdS$_5$ in Poincare coordinates as written in equation \eqref{AdS5poincare} -- however in this case $T$ is not smooth at $r=0$ and hence AdS$_5$ is not strictly within our ansatz. 

\subsection{Boundary and asymptotic conditions}

Having defined our coordinate domain, we now discuss the behaviour we impose on the metric functions at its various boundaries. Each boundary has been discussed in a general setting in the earlier \S\ref{sec:elliptic}, and we now apply those general results to our specific problem.

A defining feature of our metric is that it should be ALH with the boundary metric conformal to Euclidean Schwarzschild
\begin{equation}
h_{Schw} =  \left(1- \frac{R_0}{R} \right) d\tau^2 + \frac{dR^2}{1-\frac{R_0}{R}} +R^2 d\Omega_{(2)}^2
\end{equation}
where $\tau \sim \tau + 4\pi R_0$ is required for smoothness.
This can be written in the above coordinates by setting $R=R_0/(1-r^2)$ (so $r=0$ is the horizon) which gives
\begin{equation}
\label{Schw}
h_{Schw} = \frac{1}{f(r)^2} \left[ r^2 f(r)^2 d\tau^2 + \frac{4R_0^2 \, dr^2}{f(r)^2}+ R_0^2 \,d\Omega_{(2)}^2  \right]
\end{equation}
where $f(r)=1-r^2$ as above. It is clear from our ansatz (\ref{ansatz}) that if we impose
\begin{equation}
\label{eq:ALHbc}
A \to R_0^2\,, \qquad B \to \ell^2 \,, \qquad  T \to 1 \,, \qquad
S \to R_0^2\,,  \qquad F \to 0\,, \qquad \textrm{as} \qquad x \to 1
\end{equation}
then
\begin{equation}
\label{asymptSchw}
g_5 \sim \frac{\ell^2}{(1-x)^2 } \left( dx^2 +  \frac{f(r)^2}{4\ell^2} h_{Schw} \right)    \qquad  \qquad x \to 1
\end{equation}
which indeed shows that the conformal boundary metric is conformal to the Schwarzschild metric as required. Recall we do not assume smoothness of $X$ at $x=1$, rather the asymptotic conditions discussed in (\ref{sec:ALH}).

Following our general discussion of extremal horizons in \S\ref{sec:extremal}, 
since we wish our metric to approach the Poincare horizon (\ref{AdS5poincare}) smoothly, we must impose
\begin{eqnarray}
\label{eq:extremalbc}
&&A=\ell^2+(1-r)A_1\,, \quad B=\ell^2+ (1-r)B_1\,,     \label{extremalX1}\\
&&T=\ell^2+(1-r) T_1\,, \quad S= \ell^2+ (1-r) S_1\,, \quad F= (1-r) F_1 \,, \nonumber\\
&& T_1|_{\varrho=0} -A_1|_{\varrho=0} = \textrm{constant}  \; .  \label{extremalX2}
\end{eqnarray}
where $X_1= \{ T_1, S_1, A_1 ,B_1, F_1 \}$ are all smooth at $r=1$ and $x<1$, i.e.  the Poincare horizon. As shown in \S\ref{sec:extremal}, these conditions are necessary and sufficient for our spacetime to contain a smooth extremal Killing future horizon at $r=1$, $x<1$ which coincides with the standard Poincare horizon of AdS$_5$. 
To see this  explictly, one must pass to the Lorenzian section $t=i\tau$ and consider the coordinate transformation defined by $\varrho = 1-r^2$ and (\ref{coords:extremal1}) and (\ref{coords:extremal2}). Indeed, if the boundary conditions (\ref{extremalX1}) and (\ref{extremalX2}) are met the metric near $\varrho=0$ takes the form
\begin{eqnarray}
g_5 = \frac{\ell^2}{(1-x^2)^2} \left[ - \varrho^2 (1+O(\varrho))  dv^2 + 2 (1+O(\varrho)) dvd \varrho  +  x O(1) d\varrho dx \right. \nonumber \\ 
\left.+ \frac{4 (1+ O(\varrho) ) dx^2 } {g(x)}   + g(x) x^2 (1 +O(\varrho) ) d\Omega_{(2)}^2 \right]   
\end{eqnarray} 
thus showing $\varrho=0$ is a smooth extremal Killing horizon of $\partial / \partial v$ which coincides with that of the Poincare horizon.

It is worth mentioning at this stage that in fact (\ref{ads5NH}) is not the most general boundary condition one could take at a static extremal horizon. As is well known, the near-horizon geometry of extremal Killing horizons is in fact completely determined by the Einstein equations (see e.g. \cite{KL}). Therefore, to determine the most general boundary conditions at such a horizon in this situation one would have to provide a classification of 5d vacuum static near-horizon geometries with an $SO(3)$ symmetry. This problem has been considered in~\cite{KausReall}. It is worth noting that static 4d vacuum near-horizon geometries can be completely classified (with no assumptions of symmetry)~\cite{ChruscielReallTod} and turn out to be a 1-parameter generalisation of AdS$_4$ written in Poincare horizon adapted coordinates. The results of \cite{KausReall} indicate a 1-parameter family also in 5d. Thus, although more general static near-horizon geometries in our 5d case are likely to exist, since we are interested in the application to AdS/CFT, we restrict our attention only to the simplest case of the Poincare AdS$_5$ near horizon geometry.

\subsection{Ricci-DeTurck flow and the absence of solitons}

Our Riemannian manifold has two ends and it is ALH in one end (with conformal boundary metric in the Schwarzschild class) and asymptotically has a Lorentzian continuation to the extremal Poincare horizon of AdS in the other end.  
We must also ensure that the metric is everywhere smooth in the interior of the manifold.  Two regions where this can fail are where the $U(1)$ and $SO(3)$ symmetries have fixed points. These  correspond to $r=0$ the non-extremal horizon arising from the Schwarzschild metric on the conformal boundary, and $x=0$ the axis of the spherical symmetry, respectively. To discuss these regions properly we introduce fictitious (from the full manifold point of view) boundaries at $r=0$ and $x=0$ and work on the square $\{ 0 \leq r \leq 1, \; 0 \leq x \leq 1 \}$. 
Topologically, this square can be identified with the orbit space $\mathcal{O} \cong \mathcal{M} / [ SO(3) \times U(1) ]$, which has the natural structure of a manifold with boundaries.

As discussed in \S\ref{sec:horizon}, at the horizon $r=0$ smoothness of the manifold requires the metric functions $T,A,B,S,F$ to be smooth in $r^2$ near $r=0$, and $T = \kappa^2 A$ at $r = 0$, where $\kappa$ is a constant so that the period of $\tau$ is $2 \pi / \kappa$. The Schwarzschild conformal boundary condition given above in \eqref{eq:ALHbc} determines $\kappa = 1 /(2 R_0)$. Likewise, at the axis of spherical symmetry, we require $T,A,B,S,F$ to be smooth in $x^2$ near $x=0$, and in addition, $S/B = 1$ at $x=0$.

We consider Ricci-DeTurck flow in this class of manifolds, with $g$ given as in \eqref{ansatz} in terms of the functions $T,A,B,S,F$ with behaviours described above. We choose the reference metric $\bar{g}$ to be in the same class, and in normal coordinates to all boundaries, as we chose in the earlier \S\ref{sec:elliptic}. We make the specific choice
\begin{eqnarray}
\label{eq:g0}
\bar{g} = g \, \Big|_{T,A,B,S = 1, F = 0}   \; \; .
\end{eqnarray}
Whilst we have proved that the metric behaviour near the fictitious boundaries and the extremal horizon is preserved under the Ricci-DeTurck flow, we have no formal proof that this is so for the ALH, although it seems reasonable to expect this. We emphasize that it would be interesting to try to prove that this is so, and merely observe later in our simulations that it appears to be the case.

Suppose we flow from some initial metric in our class to a fixed point in it. We have argued in \S\ref{sec:elliptic} that no maximum of $\phi$ can occur in the ALH and extremal asymptotic regions where $\phi \rightarrow 0$, and since the other boundaries are fictitious no maximum can occur there either (as discussed, the normal derivative of $\phi$ vanishes on these fictitious boundaries). Then our earlier arguments applied here demonstrate there can be no non-trivial solitons. This is a powerful result. 
Whilst one might have worried that many solitons could in principle exist and would then complicate using Ricci-DeTurck flow to find the metric of interest, we have the beautifully simple result that in fact there can be no solitons at all.
Hence we conclude that in our case any fixed point of Ricci-DeTurck flow must be an Einstein metric, and therefore precisely relevant to the AdS/CFT problem of interest. 

\subsection{Numerical simulation of Ricci-DeTurck flow}

There are two remaining issues that could complicate using DeTurck flow to generate the Einstein metric we are interested in. 
Firstly, we have no proof that an Einstein metric satisfying the various conditions exists. Since we have ruled out the existence of solitons, this would presumably have to manifest itself in the DeTurck flow inevitably reaching a finite time singularity for \textit{any} initial data. 
Secondly, black hole fixed points have previously been shown to often be unstable under Ricci flow \cite{HeadrickTW}, due to the famous Gross-Perry-Yaffe Euclidean negative mode and its generalizations \cite{GPY}. However, in our case here the black hole metric is in part fixed at the AdS conformal boundary, and one then hopes that this projects out any putative negative modes. 
We have no way to address these questions except to simulate the DeTurck flow and observe what happens. As we shall see, we find an elegant and simple flow that quickly asymptotes to a fixed point, obviously indicating its stability and existence.

There are two scales in the problem, namely $\Lambda$ and $R_0$. We choose units so that the AdS length scale $\ell$, defined as $\Lambda = -4 / \ell^2$,  is one. Since the remaining scale $R_0$ enters only in the conformal boundary metric, and we may always rescale this since it is only the conformal class that is relevant, we may without loss of generality choose this to be one in our units.\footnote{This can be seen explicitly by defining a dimensionless angular coordinate $\theta = \tau/(2R_0)$ (which has period $2\pi$) in terms of which $h_{Schw}$ has an overall factor of $R_0^2$. Then in equation (\ref{asymptSchw}) one gets a dimensionless factor of $R_0^2/\ell^2$ in the conformal factor of the boundary metric. By then rescaling $1-x$ it is always possible to set $R_0/\ell=1$.}  Having fixed $\ell = 1$ and $R_0 = 1$, where is no remaining data in the problem. 

Given that we have a reference metric $\bar{g}$, a natural starting point for the flow is to take initial data for the flow $g(\lambda = 0) = \bar{g}$. All results presented later are for this choice. We have tried other choices, and find similar qualitative behaviour of the flow, and precisely the same fixed point.

We truncate the continuum DeTurck equations by specifying the metric functions $T,A,B,S,F$ on a finite set of points in the coordinate domain $0 \le r \le 1$, $0 \le x \le 1$. We must then implement the various behaviours on the boundaries of our domain. 
Firstly at the AdS boundary $x=1$ we simply impose the Dirichlet condition $T = A = B = S = 1$ and $F = 0$. Note our initial metric and reference metric, both $\bar{g}$ given in \eqref{eq:g0},  satisfy the ALH boundary (and fall off) conditions we defined in \S\ref{sec:ALH}. Thus from the discussion in \S\ref{parabolic:ALH}, we expect this ALH behaviour will be preserved along the flow. Indeed this is what we observe in the simulations. 
Secondly at the extremal Poincare horizon,  $r=1$ and $x<1$, we impose the Dirichlet condition $T = A = B = S = 1$ and $F = 0$. As discussed earlier, our choice of $\bar{g}$ and hence initial and reference metric asymptote smoothly to these values in $(1-r)$ and $T_1-A_1$ is a constant. Furthermore, since we proved these properties are preserved by the flow for a general static extremal horizon whose near-horizon geometry solves the Einstein equation,  we may apply this to the case at hand and deduce the flow preserves the near-horizon geometry and smoothness of the corresponding extremal horizon.

At the fictitious horizon boundary, $r=0$, the smoothness in $r^2$ implies we may simply determine the metric functions at the boundary points from their values at interior points by imposing Neumann boundary conditions on the metric functions, $\partial_r T = \ldots = \partial_r F =0$ at $r =0$. We note that the initial data and reference metric are indeed smooth in $r^2$, and satisfy the condition $T = \kappa^2 A$ at $r=0$, where $\kappa = 1/(2R_0) = 1/2$ for our choice of parameters. Since the DeTurck flow will preserve this smoothness in $r^2$ and the condition $T = \kappa^2 A$, in principle there is nothing more to do. However, we have found that directly imposing $T = \kappa^2 A$ improves numerical accuracy and stability, essentially as it eliminates potential $1/r$ divergences that arise from discretization error. 

The remaining domain boundary at $x=0$ is the fictitious axis of symmetry boundary. The required behaviour is smoothness of metric functions in $x^2$, and $S = B$ at $x=0$. This may simply be used to determine the values of the metric functions at boundary points from their values in the interior from Neumann boundary conditions, $\partial_x T = \ldots = \partial_x F =0$ at $x =0$. Again we observe that since $g_0$ is smooth in $x^2$ and obeys $S = B$ at $x=0$, the initial metric and reference metric have the desired regular behaviour for the axis of symmetry, and DeTurck flow will preserve this. As with the horizon we have found that directly imposing $S = B$ on the axis significantly improves numerical accuracy and stability.

\begin{figure}[ht]
\begin{center}
\vspace{0.5cm}
\includegraphics[scale=0.8]{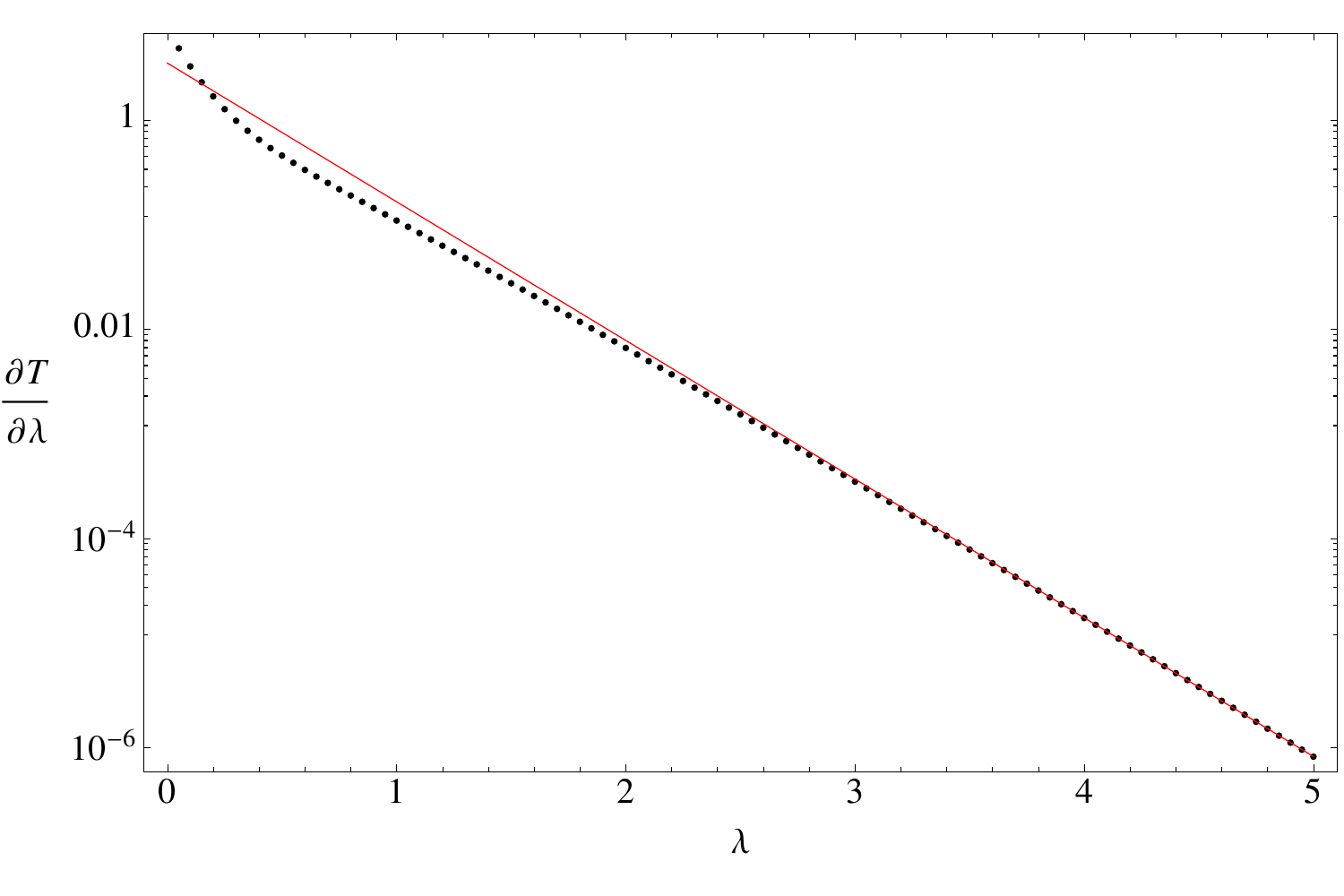}
\end{center}
\caption{Maximum value of $\frac{\partial T}{\partial \lambda}$ as a function of flow time $\lambda$ (black dots), and fit of the late time behaviour of this function to a power law $a\,\lambda^{-p-1}$ (red).}
\label{fig:evolutionT}
\end{figure}

We simulate the flow using two independent codes, both based on  forward Euler differencing in flow time; one code uses a second order finite difference in $r, x$ (up to $160 \times 160$ points), and the other a quasi-spectral representation of $r,x$ (up to $40 \times 40$ points).  
We indeed find a stable fixed point for the numerical flows, and the metric functions $T,A,B,S,F$ remain smooth in the interior of our coordinate domain, and compatible with our boundary conditions.
 In Fig. \ref{fig:evolutionT} we plot the maximum value of the flow time derivative, $\partial / \partial \lambda$, of one of the metric functions, $T$, over our coordinate domain. We see that the function
$T$ approaches the fixed point as a power law which we estimate as $\sim \lambda^{-p}$ with $p\sim2$. We see analogous scaling towards the fixed point for the remaining metric functions, and all other geometric quantities we have computed.

We are ultimately interested in the metric at the fixed point. Using Ricci-DeTurck flow to find this there are two sources of numerical error. Firstly, and most fundamentally, having spatially discretized the solution, the fixed point will solve an approximation to the Einstein-DeTurck equation, not the continuum equation. Secondly, if one simulates the flow for a finite time one will not actually reach the fixed point. 
The first source of error is inescapable, and is fundamental to any numerical approach. However, we can avoid the second source of error by making sure that we flow for long enough so that the error introduced by terminating the flow at finite flow time is small compared to the actual discretization error.

\begin{figure}[ht]
\begin{center}
\vspace{0.5cm}
\includegraphics[scale=0.5]{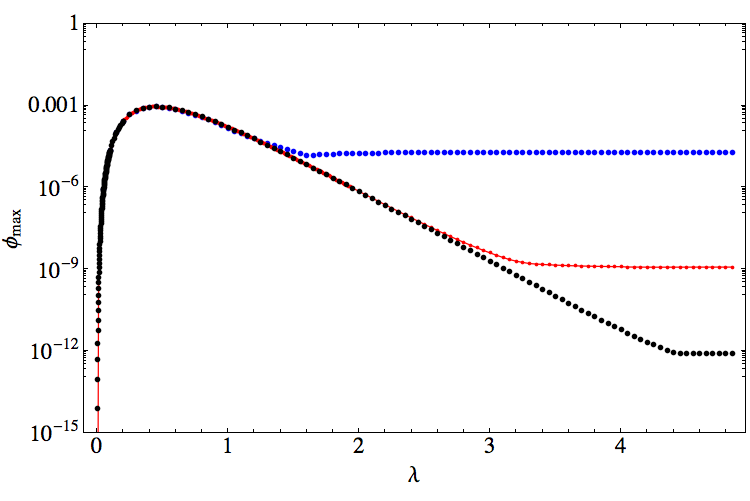}
\end{center}
\caption{Maximum value of $\phi$ in the whole domain as a function of the Ricci flow time $\lambda$ for different spatial resolutions obtained  using the quasi-spectral code.  The blue curve corresponds to the $20\times 20$ data, the red one to the $30\times 30$ data and the black curve to the $40\times 40$ data.  The value of $\phi_{\textrm{max}}$ at the fixed point (constant section of the curves above) decreases as the spatial resolution is increased. 
}
\label{fig:convergence}
\end{figure}

\begin{figure}[ht]
\begin{center}
\vspace{0.5cm}
\includegraphics[scale=0.8]{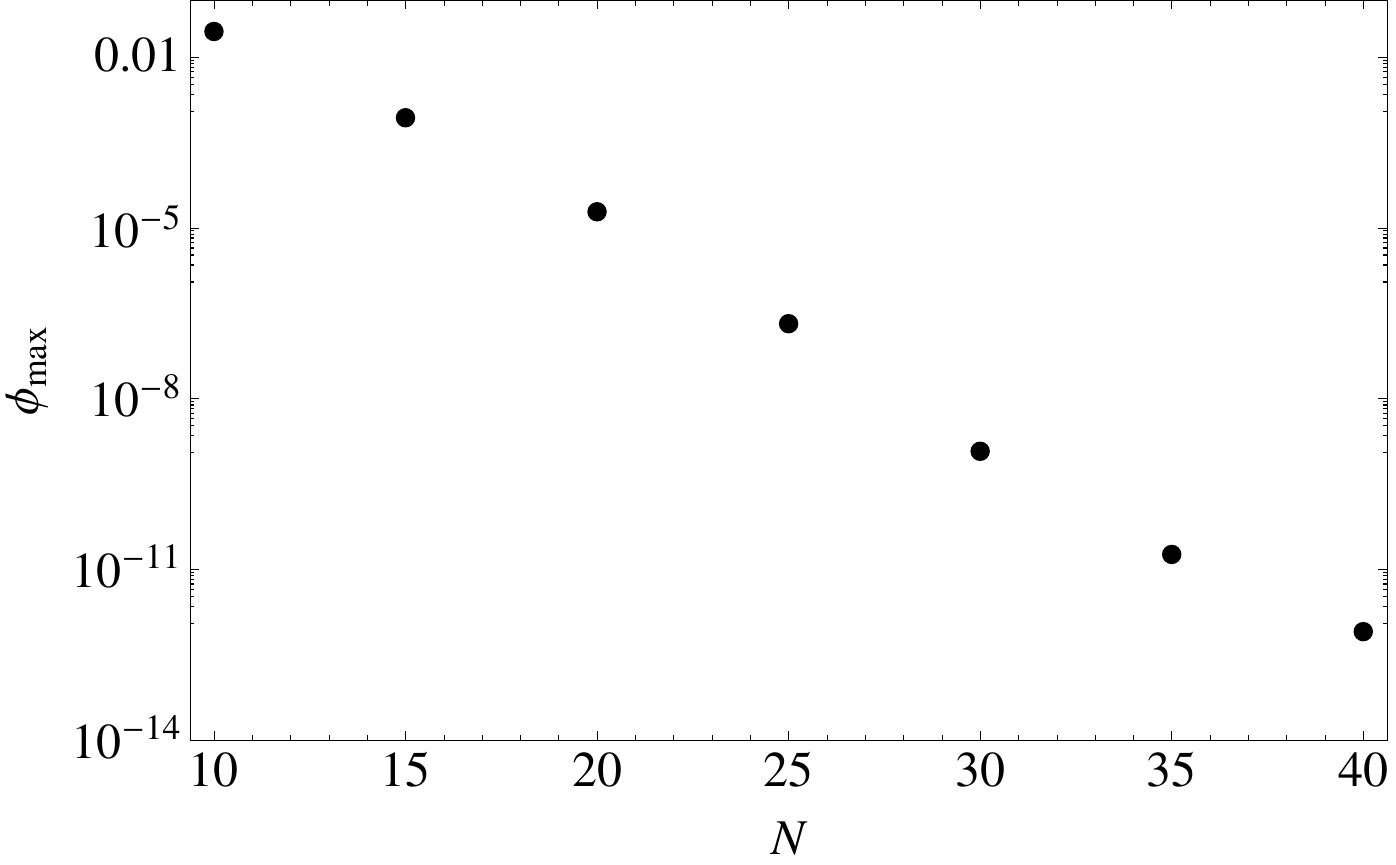}
\end{center}
\caption{Saturating value of $\phi_\textrm{max}$ as a function of the number of grid points $N$ (in either the $r$ or $x$ directions) for the quasi-spectral code. $\phi_\textrm{max}$ converges to zero exponentially with $N$, as expected for a suitably smooth solution.}
\label{fig:phimax}
\end{figure}

With our boundary conditions, the maximum principle guarantees that $\phi$ should vanish at the fixed point of the flow in the continuum, and therefore we can use this quantity to estimate the numerical discretization error for the flow. In Fig. \ref{fig:convergence} we show the evolution of the maximum value of $\phi$, denoted by $\phi_{\textrm{max}}$,  as a function of flow time. This is initially zero since we are showing results for a flow where the initial data was chosen to be the reference metric $\bar{g}$. It evolves to be non-zero, although interestingly it is never very large. At late times as we approach the fixed point in the continuum we would expect $\phi_{\textrm{max}}\rightarrow 0$ (since the fixed point must have $\xi = 0$). However, the numerical discretization error implies that at the numerical fixed point, the scalar $\phi$ will be non-zero. Taking higher resolution, the non-zero value of $\phi_{\textrm{max}}$ at the numerical fixed point should go to zero, its continuum value, scaling in accord with the nature of the discretization. The figure shows this behaviour quite clearly. Around $\lambda \sim 1.5$ the value of $\phi_{\textrm{max}}$ saturates for the $20 \times 20$ quasi-spectral flow, indicating that whilst the metric functions are still evolving as we are at finite flow time, the error in the solution is now dominated by the numerical discretization error. Increasing to $30 \times 30$ and then $40 \times 40$ resolutions we see that one must flow for longer before the finite flow error becomes subdominant to the discretization error. In Fig \ref{fig:phimax}  we plot the saturating value of $\phi_{\textrm{max}}$ as a function of resolution for the quasi-spectral flows, and find beautiful exponential convergence in the number of grid points as we should expect provided $T,A,B,S,F$ are suitably smooth functions of the coordinates. 

The square root of $\phi_{\textrm{max}}$ provides an indicator of the global fractional numerical error (since $\phi$ is contructed from the square of $\xi$), and we see that for quasi-spectral $40 \times 40$ this error is already impressively small, $< 10^{-6}$, and much smaller than our second order finite difference code achieves at the  highest resolution implemented.
Data presented from this point on will be for the quasi-spectral solution obtained by flowing for sufficient time that the numerical discretization error dominates.

In fact we have also used the Newton method, discussed in \cite{KitchenHeadrickWiseman}, to directly solve the Einstein-DeTurck equation to find the fixed point using identical spatial discretizations. We find solutions using both second order and quasi-spectral codes that are precisely consistent with those found by simulating the Ricci-DeTurck flow for long times. We emphasize that implementing the Newton method is considerably more complicated than implementing the flow method, 
although the former is considerably quicker taking hours rather than days to compute the $40\times 40$ quasi-spectral solution running in \textit{Mathematica} on a single processor machine.  However we should highlight that the path taken through the space of geometries by the Ricci-DeTurck flow method is interesting, and as emphasized in \cite{KitchenHeadrickWiseman} does not depend on the choice of reference metric. This should be contrasted with the Newton method where the iterations that improve the solution inherently depend on the choice of reference metric and therefore contain no geometric information.

\begin{figure}[ht]
\begin{center}
\includegraphics[scale=0.5]{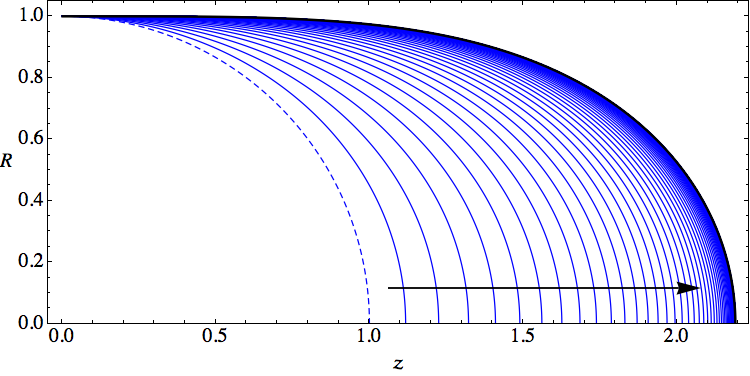}
\end{center}
\caption{Embedding into hyperbolic space, $ds^2 = \frac{\ell^2}{z^2} \left( dz^2 + dR^2 + R^2 d\Omega_{(2)}^2 \right)$, of the spatial cross sections of the horizon along the flow as curves $R(z)$. The dashed line corresponds to the initial data, for which the horizon is round, and the thick black line is the embedding of the horizon of the fixed point. The snapshots are drawn at intervals of $\lambda$ of 0.05. }
\label{fig:embed}
\end{figure}

We may embed the horizon of our geometry along the flow into hyperbolic space with the same AdS length $\ell$, in axisymmetric coordinates, ie. $ds^2 = \frac{\ell^2}{z^2} \left( dz^2 + dR^2 + R^2 d\Omega_{(2)}^2 \right)$, as the surface of revolution $R = R(z)$, such that the induced metric of this surface is the same as that of the horizon. 
The freedom in the embedding is fixed by setting $R(0) = 1$.
Fig. \ref{fig:embed} then shows the evolution of this surface from the initial data to late times in the flow. We note that the maximum extent in $z$ of the embedding approaches its fixed point value in the same manner as the metric functions discussed above, namely as $\sim \lambda^{-2}$.

We see that the shape of the horizon is very different from that of a black string, $ds^2 = \frac{\ell^2}{z^2} \left( dz^2 + h_{Schw} \right)$, where $h_{Schw}$ is the 4$d$ Schwarzschild metric (\ref{Schw}). This is also an Einstein metric, but does not have a regular Cauchy horizon, and is unstable to a Gregory-Laflamme instability \cite{GL, Gregory}.
From our embeddings we see the horizon of our solution does not look string-like for $z > 1$ (a string would have embedding $R(z) = 1$). Following the discussion in \cite{RandallTW}, one would expect a Gregory-Laflamme instability only if the embedding behaved as $R(z) \sim 1$ up to quite large $z$, say $z > 5$ before `capping off', otherwise a wavelength of the instability cannot fit into the string-like portion (see Fig. 2 in \cite{RandallTW}).
Since this is not the case, we do not expect our solution is unstable to a Gregory-Laflamme instability. 
We emphasize here that since our flow stably approaches the fixed point, our solution cannot have negative modes 
which respect the static axisymmetry.
In fact we have explicitly confirmed this by a direct calculation of the spectrum of the linearization of the Einstein-DeTurck equation about the solution assuming its symmetries.
Whilst there is no direct correlation between Euclidean negative modes and Lorentzian instability, it would have been suggestive of a potential instability had negative mode(s) existed.
Thus we find nothing to suggest dynamical instability, but emphasize that one must perform a Lorentzian linear analysis to actually determine dynamical stability. We have not done this, and it would obviously be interesting to do so.

\begin{figure}[ht]
\begin{center}
\includegraphics[scale=0.63]{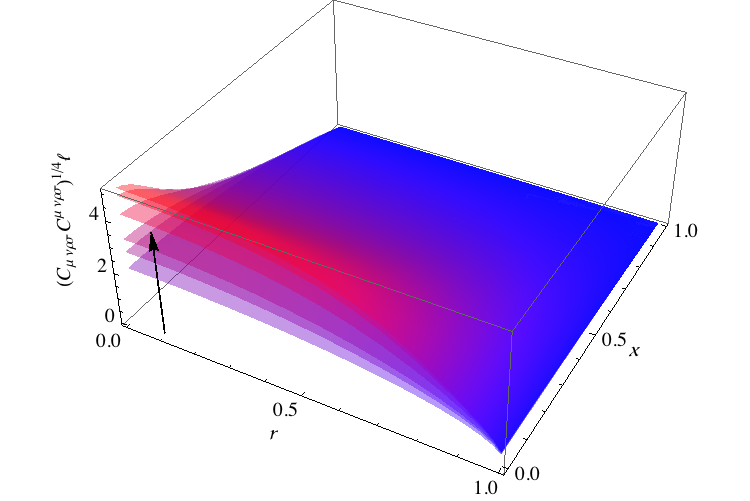}
\end{center}
\caption{Evolution of $(C_{\mu\nu\rho\sigma}C^{\mu\nu\rho\sigma})^{\frac{1}{4}}\ell$ along the flow. The snapshots correspond to $\lambda=0, 0.1,0.2, 0.5, 1$ and the fixed point. As shown in the plot, the Weyl tensor remains zero at the Poincare horizon $(r=1)$ and also at the AdS boundary $(x=1)$. The latter suggests that the ALH boundary conditions are preserved by the flow.}
\label{fig:weyl}
\end{figure}

Finally, in Fig. \ref{fig:weyl} we depict the evolution of the  curvature invariant $(C_{\mu\nu\rho\sigma}C^{\mu\nu\rho\sigma})^{\frac{1}{4}}\ell$ along the flow. As this figure shows, all along the flow the Weyl tensor vanishes at both the Poincare horizon and at the AdS boundary. 
This is consistent with the Ricci-DeTurck flow preserving ALH metrics with a fixed boundary metric, as we discussed earlier in \S\ref{parabolic:ALH}.
  At the fixed point, from the behaviour of this curvature invariant near these two asymptotic regions of the spacetime we can estimate the behaviour of the Weyl tensor, and we find that $C_{\mu\nu\rho\sigma}\sim (1-x)^2$ near the boundary of AdS and $C_{\mu\nu\rho\sigma}\sim (1-r)$ near the Poincare horizon.  We have also checked that for our solution the condition $T_1|_{r=1}-A_1|_{r=1}=\textrm{constant}$ is satisfied (within numerical accuracy). Recall from the discussion around eq. \eqref{extremalX2} that this is a 
necessary 
condition for the regularity of a static extremal Killing horizon (given the near-horizon geometry is). For the initial data (and hence our reference metric $\bar{g}$) the curvature is quite small everywhere, but along the flow it grows near the horizon and the axis of symmetry.  It is worth noting that for the fixed point solution the curvature is finite everywhere inside our coordinate domain.

\subsection{Boundary stress tensor}

Now let us consider the boundary stress tensor. For a fixed point with $\xi^\mu = 0$, one may compute the asymptotic behaviour of the functions $X$, and find,
\begin{equation}
\begin{aligned}
T &=  1 - (1-x^2)^2 (1 - r^2) + (1-x^2)^4 \,t_4(r) + \ldots \,,\\ 
S &=  1 + \frac{1}{2} (1-x^2)^2 (1 - r^2) + (1-x^2)^4\, s_4(r) + \ldots \,,\\ 
A &=  1 - (1-x^2)^2 (1 - r^2) + \frac{1}{4}(1-x^2)^4\big[3-8\,r^2+5\,r^4-4\, t_4(r)-8\,s_4(r)\big] + \ldots \,,\\ 
B  &=  1 + \frac{1}{4}(1-x^2)^4 (1 - r^4) + \ldots \,,\\ 
F  &=  - (1-x^2)^3 (1 - r^2) + \frac{3}{2} (1 - r^2)^2 ( \log(1-x^2) (1-x^2)^5 ) + f_5(r) (1-x^2)^5 + \ldots \,, 
\label{eqn:nearbdry}
\end{aligned}
\end{equation}
where $t_4(r)$ is data to be determined by fitting from the bulk solution, and $s_4(r)$ is given in terms of this from,
\begin{equation}
\label{eq:conserve}
8 (1 - 3 r^2) (s_4 + t_4 ) - ( 1 - r^2 ) \big[ 3 - 27\, r^2 + 30 \,r^4 - 4 \,r ( 2\, s_4' + t_4 ' ) \big]= 0 \, .
\end{equation}
After a coordinate transform to the usual Fefferman-Graham form
\begin{equation}
g_5 = \frac{\ell^2}{z^2}  \Big[  dz^2 + h_{Schw} + z^4 h_4 + O(z^6) \Big]
\end{equation}
with Schwarzschild conformal boundary metric  $h_{Schw}$ given by (\ref{Schw}), one can extract the leading large $N_c$ behaviour of the vev of the boundary stress tensor, $\langle T_i^{\phantom ij}\rangle$, 
\begin{equation}
\begin{aligned}
\frac{1}{N_c^2} \langle T_i^{\phantom ij}\rangle=\frac{1}{2\pi^2}\,\frac{1}{R^4}\,\textrm{diag}\bigg\{&\frac{3\,R_0}{4\,R}\left(1-\frac{R_0}{R}\right)+t_4(R)\,,\,\,\frac{3\,R_0^2}{4\,R^2}-\big(t_4(R)+2\,s_4(R)\big) \,,\\&-\frac{3\,R_0}{8\,R}+s_4(R)\,,\,\,-\frac{3\,R_0}{8\,R}+s_4(R)\bigg\}\,,
\end{aligned}
\label{eqn:stressT}
\end{equation}
where the first two terms correspond to the $\tau\tau$ and $RR$ components respectively. Equation \eqref{eq:conserve} implies the stress tensor is conserved. Notice that the stress tensor is traceless, as it should be since the boundary metric is Ricci flat and therefore there is no anomaly in the boundary CFT. 

\begin{figure}[ht]
\begin{center}
\includegraphics[scale=0.8]{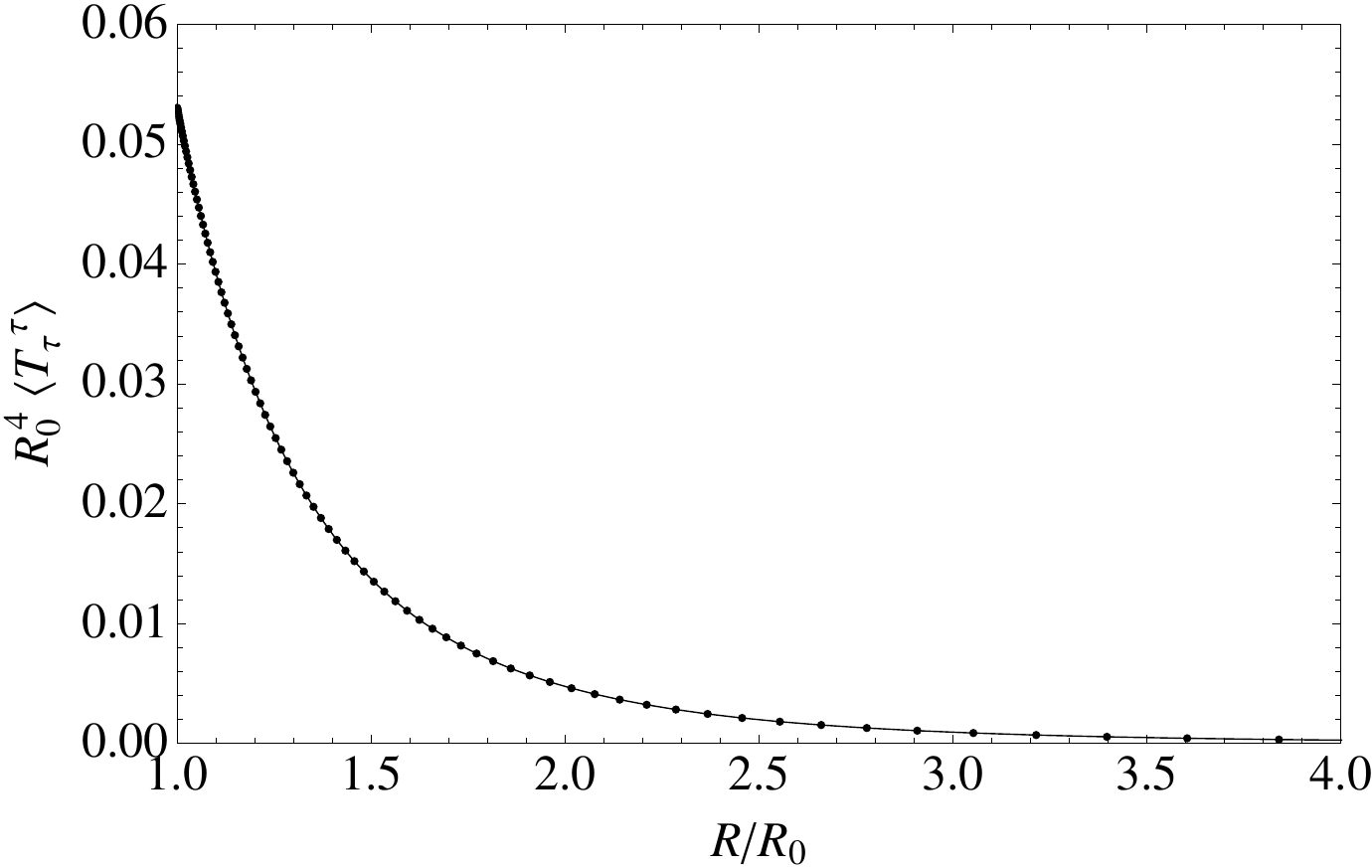}
\end{center}
\caption{Energy density divided by $N_c^2/(2\pi^2)$ as a function of the Schwarzschild radial coordinate $R$. The energy density is finite at the horizon and decays like $R^{-5}$ far from it.}
\label{fig:stressT}
\end{figure}

\begin{figure}[ht]
\begin{center}
\includegraphics[scale=0.55]{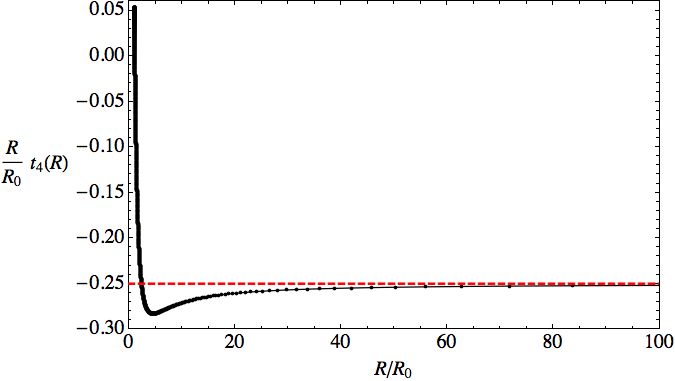}
\end{center}
\caption{$t_4(R)$ calculated from our numerical solution by fitting the near boundary region  to the behaviour in the large $R$ region, $t_4(R)$ decays as $-R_0/(4\,R)$.}
\label{fig:t4}
\end{figure}

In  Fig. \ref{fig:stressT} we plot $R_0^4\,\langle T_{\tau}^{\phantom\tau\tau} \rangle$ for the fixed point of the flow, and in Fig.  \ref{fig:t4} we show the function $t_4(R)$ that can be extracted from the numerical solution. We observe that $\frac{1}{N_c^2} \langle T_{\tau}^{\phantom \tau \tau} \rangle$ is finite at the horizon and decays like $R^{-5}$ far from the black hole.
In fact the large $R$ behaviour of the stress tensor can be conjectured analytically as follows; in the context of the Randall-Sundrum single braneworld model, the linearised gravitational field created by a point-like source (of mass $M=R_0/2$) has been calculated in \cite{Garriga:1999yh,Giddings:2000mu}. 
According to AdS/CFT from the knowledge of  the gravitational field on the brane we can compute the expectation value of the stress tensor of the CFT at strong coupling and large $N_c$ from the relation \cite{Emparan:2002px}:
\begin{equation}
G_{\mu\nu}=16\pi\,G_4\langle T_{\mu\nu}\rangle\, ,
\end{equation}
where we note that we have included an extra factor of two on the righthand side than appeared in \cite{Emparan:2002px} which using our conventions we expect to be present.
Then substituting the linear calculation of the perturbation to $G_{\mu\nu}$ for the point source yields
\begin{equation}
\frac{1}{N_c^2} \langle T_{\tau}^{\phantom\tau\tau}\rangle \underset{R\to\infty}{\approx}\frac{1}{2\pi^2}\,\frac{R_0}{2\,R^5}\,,
\end{equation}
which we observe is precisely the asymptotic behaviour of our stress tensor.

Even though the functions $t_4(R)$ and $s_4(R)$ are finite on the event horizon this is not sufficient to guarantee that \eqref{eqn:stressT} is regular there. 
However, since our bulk classical solution is a smooth geometry  in its interior we expect any geometric data extracted will also be smooth, although we note that this is not guaranteed (see for example \cite{Marolf:2010tg}).
We may explicitly check the regularity of the stress tensor by returning to Lorentzian signature and transforming  \eqref{eqn:stressT} to coordinates which are regular on the (future or past) event horizon, e.g.,  ingoing or outgoing Eddington-Finkelstein coordinates. Then the necessary and sufficient condition for regularity is that $\langle T_\tau^{\phantom \tau \tau}\rangle=\langle T_{R}^{\phantom R R}\rangle$ at $R=R_0$, but in our case this can be seen to be consequence of the conservation of the stress tensor, which in turn is implied by the bulk equations of motion. Therefore, we conclude that \eqref{eqn:stressT} is indeed regular on both the future and the past event horizon.

\section{Summary and discussion}
\label{sec:discussion}

In the first part of this paper we have shown that the existence of solutions to the Ricci soliton equation \eqref{eq:soliton} (with a non-positive cosmological constant) on an $m$-dimensional Riemannian manifold with a boundary is governed by a simple maximum principle. In particular, if the vector field $\xi$ vanishes on all boundaries, this rules out the existence of non-trivial Ricci solitons. Following \cite{KitchenHeadrickWiseman} we have considered static solutions to the Einstein-DeTurck equation by Euclidean continuation to a Riemannian problem which is elliptic. 
We have extended previous discussion by showing how to implement boundary conditions for the Einstein-DeTurck equation in a variety of cases: boundaries with `modified Dirichlet' data (Anderson's fixed trace of extrinsic curvature and conformal class for the induced metric \cite{Anderson1}), `mixed Neumann-Dirichlet'  data (extrinsic curvature proportional to induced metric), and ends where the analytically continued Lorentzian metric is asymptotically flat, Kaluza-Klein, locally AdS or has an extremal horizon. 
Furthermore, using our maximum principle, we have argued that for solutions with these boundary types, the existence of non-trivial Ricci solitons is actually ruled out.\footnote{With the exception of the mixed Neumann-Dirichlet condition for `negative tension'.} Thus if one can solve the Einstein-DeTurck equation, for example using Ricci-DeTurck flow, one is guaranteed that the solution will solve the vacuum Einstein equations (with a non-positive cosmological constant).

We also  considered whether Ricci-DeTurck flow preserves our various boundary and asymptotic conditions. We have proved that the flow does preserve the class of static asymptotically extremal manifolds -- we defined such manifolds as ones whose associated static Lorentzian spacetime contains a smooth static extremal Killing horizon whose near-horizon geometry solves the Einstein equation.  

In the second part of the paper we focussed on a particularly elegant application of these more formal results. 
We have numerically constructed a 5$d$ Einstein metric with negative cosmological term, whose static Lorentzian continuation is asymptotically AdS with boundary metric conformal to the 4$d$ Schwarzschild black hole, and whose metric far from this boundary tends to an  extremal horizon whose near horizon geometry is that of the AdS Poincare horizon. We expect this Einstein metric gives the dominant classical bulk saddle point solution for AdS-CFT where the dual strongly coupled CFT is put on a Schwarzschild background in the Unruh or Boulware vacua, so that asymptotically the theory is in vacuum. 
Our analysis of the maximum principle and boundary conditions in the first part of the paper shows not only how to formulate the elliptic Einstein-DeTurck equation and boundary conditions, but also gives the beautiful result that the maximum principle implies any solution to this DeTurck equation must be Einstein and not a Ricci soliton. 

In situations where one tries to find an Einstein solution that does not exist there are two obvious ways in which the methods above can fail. Firstly the solution found can be a soliton.\footnote{This would be the case if one tries to find a positively curved K\"ahler-Einstein metric on the second del Pezzo surface, since such a metric does not exist. A K\"ahler Ricci soliton does though, and this has indeed been found by Ricci flow  \cite{Headrick:2007fk} .} Secondly, for \textit{any} initial data Ricci flow develops a finite time singularity that may indicate a solution exists but on a manifold of a different topology.\footnote{This feature can be seen in Ricci flows relating to static black holes in cavities \cite{HeadrickTW}.} For our application it is not a priori obvious that a solution with the prescribed boundary conditions does exist, and thus ruling out the possibility of a soliton analytically is a very nice result. Of course, one can always check numerically whether one has found a soliton or not, but it is obviously preferable to have analytic control over their existence. 
If one knows solitons cannot exist, and so the DeTurck vector $\xi$ should vanish, then the norm of  $\xi$ provides a good estimator of numerical error for a solution.

Another very attractive feature of our example application is that it is particularly simple to construct the solution.
Whilst black hole solutions often have Euclidean negative modes, our solution does not, and any putative mode is projected out as the conformal boundary metric is not dynamical but is fixed to be Schwarzschild. Hence we may find the Einstein solution very straightforwardly by simulating the parabolic Ricci-DeTurck flow with some initial guess. 
With no negative modes, the solution is a stable attractor, and in principle we may flow arbitrarily close to it if we flow for sufficient time. Unlike other solutions studied with these methods, such as in \cite{KitchenHeadrickWiseman} there is no critical behaviour to be tuned in the Ricci-DeTurck flow, and no need to implement the more complicated Newton method algorithm.

We have constructed the solution to very high accuracy using pseudo-spectral methods and have exhibited various properties of the solution. In particular we have found the shape of the horizon is such that we expect no Gregory-Laflamme instability to exist. Together with the lack of Euclidean negative modes\footnote{There is of course no direct relationship between a Euclidean negative mode and a Lorentzian unstable mode as one cannot simply continue a time dependent mode and its boundary conditions between the two signatures, so a correspondence only holds for static perturbations \cite{Reall}.
That said, the existence of a Euclidean negative mode (other than that associated to local thermodynamical instability ie. a Gross-Perry-Yaffe mode, which does not occur here) might be taken as a weak indication of a potential instability.}
(at least in the static axisymmetric class as the solution is a stable fixed point of Ricci-DeTurck flow) we expect that the solution is dynamically stable, although we emphasize that a proper Lorentzian perturbation calculation would be required to determine this beyond our heuristic discussion.
In particular it is important to include the possibility of instabilities involving the internal space in AdS/CFT (for example an $S^5$) and localized near the `tip' of the bulk horizon.

We have claimed that the IR boundary conditions chosen are appropriate to describe the dominant classical bulk saddle point dual to the CFT in the Unruh and Boulware vacua. We will conclude the paper with some discussion on the interpretation of the computed CFT stress tensor.

An important point is that for any classical gravity calculation of $\langle T_{ij} \rangle$ for the dual CFT, only the leading $O(N_c^2)$ behaviour can be extracted, whilst 1-loop bulk gravity corrections would be required to compute the remaining $O(1)$ contributions to the stress tensor. 
Thus we emphasize that we have not computed the full stress tensor for the CFT but only the leading $O(N_c^2)$ part which is common to both the Unruh or Boulware vacua, these being distinguished only at $O(1)$.

Already at this leading order one can distinguish the Hartle-Hawking vacuum from the Unruh or Boulware ones, as for these latter  cases we expect $\frac{1}{N_c^2} \langle T_i^{\phantom ij}\rangle\to 0$ as $R\to\infty$, whereas for the Hartle-Hawking vacuum one expects the energy density is $O( N_c^2 )$ as $R \to \infty$. In the gravity solution, in order to describe the dominant classical bulk saddle point for the CFT in the Hartle-Hawking vacuum, one would take a finite temperature horizon in the IR, at the same temperature as the UV horizon, rather than the extremal horizon of Poincare AdS \cite{Hubeny:2009ru}. 
We note that our solution does describe \emph{some} classical bulk saddle point dual to the CFT in the Hartle-Hawking vacuum, but it is not expected to be the \emph{dominant} one.\footnote{This is analogous to considering the CFT in flat space at finite temperature where the Poincare-AdS solution does give a bulk classical saddle point but is dominated in the bulk partition function by the planar AdS-Schwarzschild saddle point.} From a Euclidean perspective this can be understood by considering which classical saddle dominates the free energy, but also from the expectation of strong corrections to the semiclassical saddle point in the extremal region due to degeneration of the Euclidean time circle which is periodically identified (with antiperiodic fermion boundary conditions) due to the finite temperature boundary black hole horizon \cite{Adams:2005rb}.

The leading $O(N_c^2)$ behaviour of our stress tensor is very interesting as since it is static, there is no energy flux to infinity and the stress tensor is regular on the past and future horizons. 
Our expectation is that by including 1-loop graviton corrections in the bulk which will determine the $O(1)$ part of the stress tensor at large $N_c$, the stress tensor will be regular only at the future horizon in the Unruh case, and will be singular on both past and future horizons in the Boulware case as is usually found.
 What makes this picture so interesting is that if the CFT were a free theory then on rather general grounds (for example \cite{Kay:1988mu}) taking the Unruh or Boulware vacua already would yield the leading $O(N_c^2)$ part of the stress tensor to be   singular on the past horizon (Unruh), or past and future horizons (Boulware), and in the Unruh case one would also see a flux of radiation in the stress tensor out to infinity.
Thus the strong interactions of the CFT appear to significantly change the behaviour of the theory from that expected by intuition from free field theory. We note that this is very much related to the issue of existence of Randall-Sundrum braneworld static black holes \cite{Tanaka:2002rb, Emparan:2002px,Emparan:1999wa,RandallTW,GregoryRoss,Hubeny:2009ru,Hubeny:2009rc}, where we believe the arguments against their existence based on extrapolation from free field theory would also incorrectly imply non-existence of the solution we have found here.

We now present a simple physical picture which we believe captures the physics of this strongly coupled CFT in the Unruh vacuum. The strong self-interaction of the CFT is attractive (corresponding to the attractive 5$d$ gravity force in the gravity dual). Thus the thermal radiation pressure from the black hole horizon must compete with the strong attractive self-interaction of the resulting plasma, which naively wishes to collapse back into the horizon of the fixed background, thus creating a ``halo" of plasma near the horizon. It appears that at order $O(N_c^2)$ in the stress tensor a static equilibrium is reached, with a thermal halo of plasma forming, and at every point in the halo the radiation pressure balances this long range attractive self-interaction.\footnote{Note that there is also the gravitational attraction to the horizon due to the background metric, but this is a small effect, and would not be expected to be strong enough to confine a thermal halo for a free CFT.} 
Although  the halo has no edge and extends all the way out to infinity, since $\frac{1}{N_c^2} \langle T_{\tau}^{\phantom\tau\tau}\rangle =  O(R^{-5})$ at large $R$, the energy of the halo is finite. While at $O(N_c^2)$ static equilibrium is reached and there is no energy flux, as we have mentioned above, the $O(1)$ component of the stress tensor will presumably exhibit Hawking radiation out to infinity in the Unruh vacuum.

It is an interesting question whether the static $O( N_c^2 )$ halo will persist to finite 't Hooft coupling, or is only an artifact of being at infinite 't Hooft coupling. In principle one might imagine studying this by including the leading $\alpha'$ corrections in the gravity dual.

Another interesting direction is to study the case of the Hartle-Hawking vacuum, which has been considered in \cite{Hubeny:2009ru,Hubeny:2009kz} where solutions are conjectured to exist and have been termed black `droplets' and `funnels'. In the case of droplets, the construction of solutions would follow essentially what we have done here, the only difference being that the extremal horizon boundary would be replaced by a non-extremal horizon `fictitious' boundary. Our maximum principle arguments would again rule out the existence of solitons, and furthermore solutions would be expected to be stable fixed points of Ricci-DeTurck flow.

\section*{Acknowledgements}

We would like to thank Michael Anderson, Roberto Emparan, Mukund Rangamani, Harvey Reall and Eric Woolgar for extremely useful discussions and comments on this work. PF is supported by an EPSRC postdoctoral fellowship [EP/H027106/1]. JL is supported by an EPSRC career acceleration fellowship. TW is supported by an STFC advanced fellowship and Halliday award.

\appendix

\section{Static extremal Killing horizons}
\label{sec:AppExtremal}

\subsection{Coordinate systems}
In this section we prove that any static spacetime containing a smooth static extremal Killing horizon, can be written in coordinates $(t,\varrho, x^a)$ valid outside the horizon $\varrho>0$, such that the metric takes the form (\ref{extremalhorizon}) where (\ref{NH}) and (\ref{smoothness}) are satisfied. Our only assumptions are that the static Killing vector field is timelike just outside the horizon and that cross-sections of the horizon are simply connected.

We begin by noting that the metric in a neighbourhood of any extremal Killing horizon of a Killing field $V$ can be written in Gaussian null coordinates $(v,r,x^a)$ as:
\begin{equation}
\label{GN}
g= 2\, dv \left(  dr+ r \, h_a(r,x) dx^a -\frac{1}{2}r^2 F(r,x) dv \right) + \gamma_{ab}(r,x) dx^a dx^b
\end{equation}
where $V=\partial / \partial v$, the horizon is $r=0$, and $x^a$ are coordinates on cross-sections of the horizon which we denote by $\mathcal{H}$. Note that in these coordinates all metric functions are smooth at $r=0$.  

We are interested in {\it static} extremal horizons, i.e.  when $V$ is hypersurface orthogonal, so $V\wedge dV=0$ everywhere. It can be shown that $V$ is hypersurface orthogonal iff
\begin{eqnarray}
&&\hat{d} h= r h \wedge \partial_r h  \label{heq} \\
&&\hat{d}F -h F +r( F \partial_r h - h \partial_rF)=0   \label{Feq}
\end{eqnarray}
where $\hat{d}$ denotes the exterior derivative at constant $r$, so e.g. $\hat{d}h= \partial_{[a} h_{b]} dx^a \wedge dx^b$, and $\partial_r h =( \partial_rh_a) \, dx^a$.  By smoothness we may write
\begin{equation}
\label{rexpansion}
F= F_0(x) +r F_1(x) + O(r^2) \;,\qquad h = h_0(x) +r h_1( x) +O(r^2)
\end{equation}
where  $F_0,F_1, h_0,h_1$ are smooth on $\mathcal{H}$. Substituting these into (\ref{heq}) and (\ref{Feq}) gives
\begin{eqnarray}
&&\hat{d} h_0 =0 \;, \qquad \qquad \qquad  \hat{d}F_0 =F_0 h_0  \label{0eq}\;, \\ 
&&\hat{d} h_1 = h_0 \wedge h_1 \;,\qquad \qquad \hat{d}F_1 = 2F_1 h_0  \label{1eq}  \; .
\end{eqnarray}
The two equations (\ref{0eq}) are easily solved
\begin{equation}
h_0 = d\lambda(x) \;,\qquad \qquad \qquad F_0=c_0 e^{\lambda(x)}  \label{F0}
\end{equation}
where $c_0$ is a constant and $\lambda(x)$ a function on $\mathcal{H}$, which agrees with the general form for a static near-horizon geometry found in~\cite{KLR}. For simply connected $\mathcal{H}$ these expressions are valid globally. The assumption that $V$ is timelike just outside the horizon, implies $c_0>0$, and w.l.o.g. we will set  $c_0=1$. The near-horizon geometry can then be written as (\ref{NHG}) where $\varrho = e^\lambda r$ and $T_0=e^{-\lambda}$.  We are now interested in generalising this to the full spacetime metric.  For later convenience we note that we can solve (\ref{1eq})
\begin{eqnarray}
h_1= e^{\lambda(x)}\,  d \lambda_1(x) \;, \qquad  \qquad \qquad F_1= c_1 e^{2\lambda(x)}   \label{F1}
\end{eqnarray}
where $\lambda_1(x)$ is a function on $\mathcal{H}$ and $c_1$ a constant. 

It is clear one could proceed by solving (\ref{heq}) and (\ref{Feq}) order by order in $r$. Instead we now find new coordinates which make the static isometry manifest.  Assuming $r>0$ and $F>0$ (as is the case if $V$ is timelike just outside the horizon), it can be shown that $V\wedge dV=0$ implies there is a coordinate transformation defined by
\begin{equation}
t=v-f(r,x) \;,\qquad \qquad df= \frac{dr+rh}{r^2F}
\end{equation}
such that the spacetime metric (\ref{GN}) becomes manifestly static
\begin{eqnarray}
g &=& -r^2 F dt^2 + r^2 F df^2 +\gamma_{ab}dx^a dx^b \nonumber  \\
 &=&  -r^2 F dt^2 + \frac{1}{F} \left( \frac{dr}{r} + h_a dx^a \right)^2 +\gamma_{ab}dx^a dx^b   \; .
\end{eqnarray}
It is convenient to change radial coordinate aswell. Thus define
\begin{equation}
\varrho =\Gamma r 
\end{equation}
for some positive smooth function $\Gamma$ chosen such that $\Gamma|_{\varrho=0}= e^{\lambda(x)} $. The metric then becomes
\begin{equation}
\label{statextreme}
g=- \varrho^2 T dt^2 +  R \left( \frac{d\varrho}{\varrho} + \varrho\,  \omega_a dx^a \right)^2 +\gamma_{ab}dx^a dx^b
\end{equation}
where
\begin{equation}
T\equiv \frac{F}{\Gamma^2} \;, \qquad R \equiv \frac{1}{F} \left( 1- \frac{\varrho\, \partial_\varrho \Gamma}{\Gamma}  \right)^2 \;,\quad
h_a \equiv  \frac{\partial_a \Gamma}{\Gamma} + \varrho \left( 1- \frac{\varrho\, \partial_\varrho \Gamma}{\Gamma}  \right) \omega_a\;,
  \label{omega}
\end{equation}
define the smooth functions $T,R$ and 1-form $\omega_a$. Note that as defined  in equation (\ref{omega}), $\omega_a$ is smooth at $\varrho=0$ as a consequence of the fact that $h_0 =d\lambda$ and the choice $\Gamma_0=e^{\lambda(x)}$.

We have thus derived the form of the metric claimed in the main text (\ref{extremalhorizon}). Furthermore, since we have related this to Gaussian null coordinates we may now deduce conditions on $T,R,\omega_a,\gamma_{ab}$ which are necessary and sufficient for the existence of a smooth static extremal horizon. First it is clear that $T|_{\varrho=0} =R|_{\varrho}= e^{-\lambda}$ which confirms (\ref{NH}).  Next, defining $T_1=\partial_\varrho T|_{\varrho=0}$ and $R_1=\partial_\varrho R|_{\varrho=0}$, one can show that
\begin{equation}
T_1-R_1= \frac{2 F_1 e^{-\lambda}}{F_0^2}
\end{equation}
where $F_0,F_1$ are defined by (\ref{rexpansion}). Now, using the equations for $F_0,F_1$ worked out above (\ref{F0}) and (\ref{F1}) implies
\begin{equation}
\frac{T_1-R_1}{T_0}= 2c_1
\end{equation}
which confirms (\ref{smoothness}).  In the main text it was shown that given the metric (\ref{statextreme}),  these conditions are sufficient to guarantee a smooth extremal horizon at $\varrho=0$.

This completes the proof that outside {\it any} smooth static extremal Killing horizon, the metric can be written as (\ref{extremalhorizon}), where $T,R,\omega_a,\gamma_{ab}$ are smooth functions at $\varrho=0$ together with (\ref{NH}) and (\ref{smoothness}),  {\it assuming} the static Killing field is timelike just outside the horizon and $\mathcal{H}$ is simply connected.

\subsection{Ricci-DeTurck flow of asymptotically extremal manifolds}
In this section we show that Ricci-DeTurck flow preserves static asymptotically extremal Riemannian manifolds. These were defined in the main text and are equivalent to the class of Lorentzian spacetimes containing a smooth extremal static Killing horizon. As argued, we may choose coordinates such that near the horizon this class of metrics can be written as (\ref{extremalend}). We already argued that the near-horizon geometry is preserved, simply as a consequence of the fact it is a fixed point of the flow equation. Now, we derive flow equations for the ``first-order" quantities $T_1= \partial_{\varrho}T|_{\varrho=0}$ and $R_1= \partial_{\varrho} R|_{\varrho=0}$. The obvious thing to do is calculate the Ricci tensor for the full metric (\ref{extremalend}) and Taylor expanding to first order. However, this is somewhat cumbersome, and we will take a different approach. Instead we will treat the first order terms as a linearised perturbation about the near-horizon geometry. 

Set $\varrho= \epsilon r$ and $\tau=t / \epsilon$ for $\epsilon>0$. The full metric (\ref{extremalend}) is then:
\begin{equation}
g_\epsilon= T_{\epsilon} r^2 dt^2 + R_\epsilon \left(  \frac{dr}{r} + \epsilon r \omega^{\epsilon}_a dx^a \right)^2 + \gamma^\epsilon_{ab} dx^a dx^b
\end{equation}
where $T_\epsilon= T(\epsilon r, x)$ etc. This is a 1-parameter family of metrics which coincide with $g_{NH}$ when $\epsilon \to 0$, i.e. $g_\epsilon \to g_0=g_{NH}$ as $\epsilon \to 0$. Now define the first order perturbation by
\begin{equation}
h =\left. \frac{dg_{\epsilon}}{d\epsilon}  \right|_{\epsilon=0}  \; .
\end{equation}
Similarly one can define the 1-parameter family of reference metrics $\bar{g}_\epsilon$, which also satisfies $\bar{g}_0=g_{NH}$, and $\bar{h}= (d\bar{g}/d\epsilon)_{\epsilon=0}$.
Differentiating the full flow equation $\partial_\lambda g_{\mu\nu} = -2( R_{\mu\nu} -\nabla_{(\mu} \xi_{\nu)} -\Lambda g_{\mu\nu})$ wrt $\epsilon$ implies a flow equation for $h_{\mu\nu}$\footnote{We have used $\dot{R}_{\mu\nu}|_{\epsilon=0} = \frac{1}{2}\Delta_L h_{\mu\nu} + \nabla_{(\mu } v_{\nu)}$ where $v_{\mu}= \nabla^\nu h_{\mu\nu}-\frac{1}{2} \partial_\mu h$, and $\frac{d}{d\epsilon}\left(2\nabla_{(\mu} \xi_{\nu)} \right)_{\epsilon=0} = \mathcal{L}_{\xi_0} h_{\mu\nu} + 2\nabla_{(\mu} \dot{\xi}_{\nu)}|_{\epsilon=0} $, and $\xi_0=\xi_{\epsilon=0}=0$ and $\dot{\xi}_{\mu}|_{\epsilon=0}=  v_\mu-\bar{v}_\mu$. Note we are using $g_0=\bar{g}_0$.}
\begin{equation}
\label{hflow}
\frac{\partial h_{\mu\nu}}{\partial \lambda} = - \Delta_L h_{\mu\nu} -2\nabla_{(\mu} \bar{v}_{\nu)}+2\Lambda h_{\mu\nu}
\end{equation}
where $\bar{v}_{\mu} =\nabla^\nu \bar{h}_{\nu \mu}-\frac{1}{2} \partial_\mu \bar{h}$, $\Delta_L$ is the Lichnerowicz operator, and all quantities are with respect to the ``background" metric $g_0=g_{NH}$ (\ref{NHG}).  Explicitly, the Lichnerowicz operator is given by
\begin{equation}
\Delta_L h_{\mu\nu} = -\nabla^2 h_{\mu\nu} + 2R_{\rho ( \mu} h_{\nu )}^{~~ \rho} -2R_{\mu \rho \nu \sigma} h^{\rho \sigma}    
\end{equation} 
where $\nabla$ and $R_{\mu\nu}$ and $R_{\mu\nu\rho\sigma}$ are the Levi-Civita connection, the Ricci tensor and Riemann tensor of $g_{NH}$.
Therefore all curvature calculations are with respect to $g_{NH}$ (\ref{NHG}), making the task considerably simpler.  In our case one can check
\begin{equation}
h= T_1 r^3 dt^2 + \frac{R_1 dr^2}{r} + 2T_0 \omega_a^1 dx^a dr + r\gamma^1_{ab} dx^a dx^b
\end{equation}
where $T_1=  \partial_\varrho T |_{\varrho=0}$, $R_1=\partial_{\varrho} R|_{\varrho=0}$, $\omega^1_a= \omega_a|_{\varrho=0}$ and $\gamma^1_{ab}= \partial_{\varrho} \gamma_{ab}|_{\varrho=0}$ all depend only on $x$. Since we are interested in obtaining an evolution equation for $T_1-R_1$  we need only consider the $tt$ and $rr$ components of (\ref{hflow}). 

Working in the coordinate basis defined by $x^\mu= (t,r,x^a)$ gives the following non-zero Christoffel symbols for the metric $g_{NH}$ (\ref{NHG}):
\begin{equation}
\begin{aligned}
&r^{-2}\Gamma^{r}_{tt}=r^2\Gamma^r_{rr}=-r \;,\qquad r^{-2}\Gamma^a_{tt}=r^2\Gamma^a_{rr}= -\frac{1}{2} \gamma^{ab} \partial_b T_0\;,
 \\
&\Gamma^t_{tr}= \frac{1}{r} \;, \qquad \qquad \Gamma^t_{ta}=\Gamma^{r}_{ra} = \frac{\partial_aT_0}{2T_0} \;, \qquad  \qquad \Gamma^a_{bc}= \Gamma[\gamma^0]^a_{bc}  \; .
\end{aligned}
\end{equation}
The relevant parts of the curvature tensors are then
\begin{eqnarray}
&& R_{trtr} = -T_0 -\frac{(dT_0)^2 }{4}  \;,\quad r^{-2}R_{tatb} = r^2 R_{rarb}= \frac{1}{2} \left( - \nabla_b \partial_a T_0 + \frac{\partial_a T_0 \partial_b T_0}{2T_0} \right) \;,    \\
&& r^{-2} R_{tt} = r^2 R_{rr} = -\frac{1}{2}\nabla^2 T_0 -1  
\end{eqnarray}
where $\nabla$ and $\cdot$ now refer to the Levi-Civita connection and contraction with respect to the metric $\gamma^0_{ab}$, respectively.
It can then be shown that
\begin{eqnarray}
&&\Delta_L h_{tt} = r^3 \left[ - \nabla^2 T_1 +\frac{dT_0 \cdot dT_1}{T_0} -\frac{2T_1}{T_0} - \frac{(T_1-R_1) (dT_0)^2}{2T_0^2} \right. \nonumber \\
&& \qquad \qquad \qquad \qquad \left. -2dT_0 \cdot \omega^1+ \gamma_1^{ab} \left( \nabla_a \partial_b T_0 - \frac{\partial_aT_0 \partial_b T_0}{T_0} \right) \right]  \\  &&\Delta_L h_{rr} = \frac{1}{r} \left[ -\nabla^2 R_1 + \frac{ dT_0 \cdot dR_1}{T_0} - \frac{2R_1}{T_0} +\frac{(T_1-R_1) (dT_0)^2}{2T_0^2}  \right. \nonumber \\ && \left. \qquad \qquad \qquad \qquad -2 dT_0 \cdot \omega^1 + \gamma_1^{ab} \left(  \nabla_a \partial_b T_0 - \frac{\partial_aT_0 \partial_b T_0}{T_0} \right) \right]    \; .
\end{eqnarray}
Therefore we deduce
\begin{equation}
\frac{1}{r^3} \Delta_L h_{tt} -r \Delta_L h_{rr}= -\nabla^2 (T_1-R_1)+ \frac{dT_0}{T_0} \cdot d(T_1-R_1) -\frac{2(T_1-R_1)}{T_0}-  \frac{(T_1-R_1) (dT_0)^2}{T_0^2}  \; .
\end{equation}
We also need $\nabla_t \bar v_t$ and $\nabla_r \bar{v}_r$. It is easily checked that $\bar{v}_t=0$ and $\partial_r \bar{v}_r=0$. This is sufficient to show that
\begin{equation}
\nabla_t \bar{v}_t = r^3 \bar{v}_r +\frac{r^2 \gamma^{ab}\bar{v}_a \partial_b T_0}{2}  \;, \qquad \nabla_r \bar{v}_r = \frac{1}{r} \bar{v}_r + \frac{\gamma^{ab}\bar{v}_a \partial_b T_0}{2r^2}
\end{equation}
and therefore
\begin{equation}
\frac{1}{r^3} \nabla_t \bar{v}_t - r \nabla_r \bar{v}_r = 0  \; .
\end{equation}
Putting all this together, we have the flow equation
\begin{equation}
\label{T1R1flow}
\frac{\partial}{\partial \lambda} (T_1-R_1) = \nabla^2 (T_1-R_1) - \frac{dT_0}{T_0} \cdot d(T_1-R_1) +(T_1-R_1) \left( \frac{(dT_0)^2}{T_0^2} +\frac{2}{T_0} +2\Lambda \right)  \; .
\end{equation}
Now, define the function
\begin{equation}
\psi \equiv \frac{T_1-R_1}{T_0}   \; .
\end{equation}
Note the $tt$ component of the near-horizon geometry Einstein equation reduces to
\begin{equation}
\nabla^2 T_0 +2 +2\Lambda T_0 =0 \; .
\end{equation}
Combining this with the flow equation for $T_1-R_1$ (\ref{T1R1flow}) one can derive a remarkably simple flow equation for $\psi$:
\begin{equation}
\frac{\partial \psi}{\partial \lambda} = \nabla^2 \psi + \frac{dT_0}{T_0} \cdot d\psi   \; .
\end{equation} 
This shows that if $\psi$ is constant at $\lambda=0$, then by uniqueness of the flow equation it must be constant for $\lambda>0$.  This is sufficient to establish that asymptotically extremal manifolds are preserved by Ricci-DeTurck flow, provided the near-horizon geometry solves the Einstein equations.

\section{Static non-extremal Killing horizons}
\label{sec:AppNonExtremal}

Consider any static spacetime containing a smooth, static, non-extremal Killing horizon of $V$, such that  $V$ is timelike just outside the horizon. In this section we prove that outside the horizon the metric can always be written in coordinates $(t,w, x^a)$ such that the metric takes the form (\ref{nonextremehorizon}) for $w>0$ where $w=0$ is the horizon. 

We begin by noting that the metric in a neighbourhood of any Killing horizon of a Killing field $V$ can be written in Gaussian null coordinates $(v,r,x^a)$ as:
\begin{equation}
\label{GN2}
g= 2\, dv \left(  dr+ r \, h_a(r,x) dx^a -\frac{1}{2}r f(r,x) dv \right) + \gamma_{ab}(r,x) dx^a dx^b
\end{equation}
where $V=\partial / \partial v$, the horizon is $r=0$, and $x^a$ are coordinates on cross-sections of the horizon which we denote by $\mathcal{H}$. Note that in these coordinates all metric functions are smooth at $r=0$.  

We are interested in {\it static} horizons, i.e.  when $V$ is hypersurface orthogonal, so $V\wedge dV=0$ everywhere. By smoothness we can write $f=f_0(x)+O(r)$ for some smooth function $f_0$ on $\mathcal{H}$ (and similarly for $h_a$ and $\gamma_{ab}$). Expanding the $vrx^a$ component of the staticity condition $V\wedge dV=0$ around $r=0$, then allows one to show that $\partial_af_0=0$. Therefore
\begin{equation}
\label{kappa}
f_0 = 2 \kappa
\end{equation}
where $\kappa$ is a constant which is easily shown to be the surface gravity. If $\kappa=0$ then the horizon is extremal, and therefore in this section we assume $\kappa \neq 0$. Since we assume $V$ is timelike just outside the horizon we deduce $\kappa>0$.

Staticity implies that outside the horizon $r>0$, there exists a coordinate transformation defined by 
\begin{equation}
t= v- \lambda\;, \qquad \qquad d\lambda = \frac{dr+ rh}{rf}
\end{equation}
in terms of which the metric becomes
\begin{equation}
g= -rf dt^2 +\frac{1}{rf} ( dr + r h_a dx^a)^2 +\gamma_{ab}dx^a dx^b   \; .
\end{equation}
Now, for $r>0$, change the radial variables by defining
\begin{equation}
w^2 = \Gamma(r,x) r
\end{equation}
for some positive smooth function $\Gamma$. Then, for $w>0$, the metric becomes
\begin{equation}
\label{nonextremehorizon2}
g= - w^2 T^2 dt^2 + W^2 (dw + w \Omega_a dx^a )^2 +\gamma_{ab}dx^a dx^b
\end{equation}
where we have defined
\begin{equation}
T^2 \equiv \frac{f}{\Gamma} \;, \qquad W^2 = \frac{4}{f\Gamma} \left( 1- \frac{w \partial_w\Gamma}{\Gamma} \right)^2   \;,\qquad
 \Omega_a = \frac{h_a - \frac{2\partial_a\Gamma}{\Gamma}}{2\left(1- \frac{w \partial_w\Gamma}{\Gamma}\right) }\;,
\end{equation}
where $T,W>0$, which is indeed of the form claimed in the main text (\ref{nonextremehorizon}). Notice that we have used the assumption that $f>0$ near $r>0$, i.e. $V$ is timelike just outside the horizon, to define $T,W$.

We may now deduce necessary and sufficient conditions on the metric functions $T, W, \Omega_a, \gamma_{ab}$  for (\ref{nonextremehorizon2}) to describe a smooth non-extremal horizon at $w=0$. First note that in the original coordinate $r$ these metric functions are smooth at $r=0$, and hence we deduce they must be smooth in $w^2$ (not $w$) at $w=0$. Also we have
\begin{equation}
\left. \frac{T^2}{W^2} \right|_{w=0} =\left.  \frac{f^2}{4} \right|_{r=0}= \kappa^2
\end{equation}
where in the last equality we used (\ref{kappa}).  As we show in \S\ref{sec:horizon}, these conditions are sufficient to ensure that the metric (\ref{nonextremehorizon2}) describes a smooth non-extremal horizon at $w=0$. This completes the proof of the claim in \S\ref{sec:horizon}.

\section{A comment on Ricci flow in ALH manifolds}
\label{sec:AppALH}

The idea of the Ricci-DeTurck flow, and indeed the Einstein-DeTurck equation is that one can avoid choosing an explicit coordinate system in order to gauge fix the Einstein equations. Rather the DeTurck vector provides extra degrees of freedom to ensure the equations are parabolic and elliptic respectively.  However, for Ricci flow in ALH manifolds, an obvious question is whether the Fefferman-Graham (FG) expansion is valid. Although somewhat outside the scope of this paper, for completeness we present a simple argument which indicates that this is not the case.

Consider the scalar diffusion equation
\begin{equation}
\partial_\lambda \phi = \nabla^2 \phi
\end{equation}
in the $n$-dimensional hyperbolic background
\begin{equation}
g= \ell^2 \left[  \frac{d\rho^2}{4\rho^2} + \frac{h_{ij}(x)dx^idx^j}{\rho} \right]
\end{equation} 
which satisfies $\text{Ric}(g)=-(n-1)\ell^{-2} g$ iff $\text{Ric}(h)=0$. In these coordinates the defining function $z^2=\rho$.  It is easy to show that explicitly it is given by
\begin{equation}
\label{flowphi}
\dot{\phi}= \frac{\rho}{\ell^2} [ -2(n-3) \phi' +\hat{\nabla}^2 \phi ] + \frac{4\rho^2 \phi''}{\ell^2}
\end{equation}
where $\dot{\phi}=\partial_\lambda \phi$ and $\phi'=\partial_\rho \phi$ and $\hat{\nabla}$ is the connection of $h$. 

For simplicity we will consider only the case of even dimension $n$. Then we may try to find a Frobenius expansion for the scalar flow as one does for the case of Laplace's equation. Such an expansion will have the form,
\begin{eqnarray}
\label{eq:exp}
\phi(\rho,x)= \sum_{k \geq 0} \left( \phi_{2 k}(x) \rho^k + {\phi}_{2 k + n - 1}(x) \rho^{(n - 1)/2 + k} \right)
\end{eqnarray}
and we note in odd dimensions one can perform a similar expansion but require logarithmic terms. Expanding and solving (\ref{flowphi}) order by order in $\rho$ gives the following relations:
\begin{equation}
\begin{aligned}
&\dot{\phi}_0=0 \;,\qquad \qquad \dot{\phi}_{n-1}=0\;,  \\
& \dot{\phi}_{m}= -\frac{m}{\ell^2} ( n-1-m) \phi_{m} + \frac{1}{\ell^2} \hat{\nabla}^2 \phi_{m-2} \;, \qquad \mathrm{for} \quad \left\{ \begin{array}{c} m > 0 \, , \, m \; \mathrm{even} \\ m > n-1 \, , \, m \; \mathrm{odd} \end{array} \right.  
\end{aligned}
\end{equation}
We see that the two terms in the expansion \eqref{eq:exp} are not coupled together by the recursion relation. Consider first the integer powers of $\rho$. The boundary value of $\phi$ (the `non-normalization data' in AdS-CFT language) given by $\phi_0$, does not evolve and we may solve for $\phi_2$ as a function of $\lambda$:
\begin{equation}
\phi_2= \frac{ \hat{\nabla}^2 \phi_0 }{2(n-3)} +  u_2 e^{-\frac{2(n-3) \lambda}{\ell^2}}
\end{equation}
where $u_2$ is independent of $\lambda$. For $2k < n-1$ we can recursively  solve for $\phi_{2k}$ and find that 
\begin{equation}
\phi_{2k} = \frac{\hat{\nabla}^2 \phi_{2(k-1)} }{2k(n-1-2k)} + \dots
\end{equation}
where $\dots$ represent terms which decay exponentially in $\lambda$, and therefore $\phi_{2k}$ converges to a value determined by $\phi_0$. Then $\phi_{n-2}$ is the last coefficient which decays exponentially in $\lambda$, and for $2k \geq n$ we see that $\phi_{2k}$ are still determined however they grow exponentially in $\lambda$ and hence do not converge at late flow time. For the half integer powers the leading term $ \phi_{n-1}$  (the `normalization data')  does not evolve. However all the higher odd terms $\phi_{2k + 1}$ grow exponentially in $\lambda$, and so again do not converge.

This example shows that the small $\rho$ expansion \eqref{eq:exp} for $\lambda>0$ breaks down at $O(\rho^{1+(n-1)/2})$, since for large $m$ one gets terms in the $\rho$ expansion which grow exponentially as $\sim \exp( m^2 \lambda /\ell^2) \rho^{m/2}$. One obtains a similar result for odd dimensions.

We expect analogous  results for the Ricci flow thus indicating that the FG expansion in $\rho$ cannot be used to describe the flow in some constant neighbourhood of the boundary. This means that there it is unclear how to define the boundary stress tensor along the flow.  
We note though that, curiously, the lower order terms $2k < n-1$ in the expansion all converge exponentially to the values at a fixed point (where the FG expansion is valid), just as in the example of the scalar field above.

\bibliographystyle{JHEP}
\bibliography{flow}

\end{document}